\documentstyle[aaspp4,epsf,graphics]{article}
\begin{document}


\def\lesssim{\mathrel{\hbox{\rlap{\hbox{\lower4pt\hbox{$\sim$}}}\hbox{$<$}}}}
\def\gtrsim{\mathrel{\hbox{\rlap{\hbox{\lower4pt\hbox{$\sim$}}}\hbox{$>$}}}}
\def \etal      {{\it et al.}\ }
\def \se       {\!=\!}
\def \sequiv   {\! \equiv \!}
\def \Ho        {{\rm H_{o}}}
\def \kmsmpc    {{\rm\ km\ s^{-1}\ Mpc^{-1}}}
\def \kev       {{\rm\ keV}}
\def \deltac    {\Delta}

\title{Four Measures of the Intracluster Medium Temperature and
their Relation to a Cluster's Dynamical State}

\author{B. F. Mathiesen} \affil{Dept. of Physics,
Stanford University, Stanford, CA 94305 USA}
\author{A. E. Evrard} \affil{Dept. of Physics,
University of Michigan, Ann Arbor, MI 48109-1120 USA}

\begin{abstract}
We employ an ensemble of 24 hydrodynamic cluster simulations to create
spatially and spectrally resolved images of quality comparable to {\em
Chandra}'s expected performance.  Emission from simulation mass
elements is represented using the XSPEC {\texttt mekal} program
assuming 0.3 solar metallicity and the resulting spectra are fit with
a single-temperature model.  Despite significant departures from
isothermality in the cluster gas, single-temperature models produce
acceptable fits to 20,000 source photon spectra.  The spectral fit
temperature $T_s$ is generally lower than the mass weighted average
temperature $T_m$ due to the influence of soft line emission
from cooler gas being accreted as part of the hierarchical
clustering process. 

The nature of this deviation depends on the bandpass used for
spectral fitting.  In a {\em Chandra}-like bandpass of 0.5
to 9.5 keV we find a nearly uniform fractional bias of
$(T_m-T_s)/T_s \simeq 20\%$, although smaller clusters sometimes
demonstrate much greater deviations.  If the minimum energy threshold
is raised to 2 keV, however, the effect of line emission on
the spectrum is greatly decreased and $T_s$ becomes a 
nearly unbiased estimator of $T_m$ for smaller clusters. The fractional
deviation in $T_s$ relative to $T_m$ is scale-dependent in this
bandpass and follows the approximate relation
$(T_m-T_s)/T_s = 0.2\log_{10}T_m$.
This results in an observed $M_{ICM}$--$T_s$ relationship for
the simulations with slope of about 1.6, intermediate between the
virial relation $M \propto T_m^{3/2}$ and the observed relation
$M_{\rm ICM} \propto T^2$.

Tracking each cluster in the ensemble at 16 epochs in its evolutionary
history, we catalogue merger events with mass ratios exceeding 10\%
in order to investigate the relationship between spectral temperature and
proximity to a major merger event.  Clusters that are very cool
relative to the mean mass-temperature relationship lie preferentially
close to a major merger, suggesting a viable observational method to
cull a subset of dynamically young clusters from the general population.

\end{abstract}

\keywords{cosmology: observations --- galaxies: clusters: general ---
intergalactic medium --- X--Rays: general}

\section{Introduction}

Galaxy clusters are the youngest and largest organized structures in
the universe, and as such provide us with a wealth of cosmological
information.  The most massive clusters draw their substance from
cosmologically significant volumes of linear scale $\gtrsim 10 
h^{-1}$ Mpc.  These scales are large enough that no known coherent
process competes against gravity, so rich cluster contents are thought
to comprise a fair sample of the universe's ingredients (White \etal
1993).  Because clusters are rare nonlinear excursions of the cosmic
density field, the statistical properties of their population are
quite sensitive to both cosmological model and slope of the primordial
fluctuation spectrum. Unfortunately, their relative youth
can also make interesting physical properties difficult to measure:
about\ 50\% of the local population bears evidence of ongoing mergers,
and the canonical ``relaxed'' cluster is a relatively rare beast.

Nearly all interesting cosmological tests depend on accurate
measurement of cluster virial masses.  Observations of the intracluster
medium (ICM) have shown promise in this regard: the ICM's high X-ray
luminosity and large spatial extent make it possible to probe
the content and structure of clusters in great detail.  
Bulk properties of the ICM such as luminosity, temperature, mass, 
and gas density profile shape have been found
to display highly significant correlations with each other 
(Edge \& Stewart 1991; David \etal 1993; Mohr \& Evrard 1997;
Mushotzky \& Scharf 1997; Markevitch 1998; Allen \& Fabian 1998;
Mohr, Mathiesen \& Evrard 1999; Arnaud \& Evrard 1999).  In contrast
to the noisy correlations of early X-ray data (Sarazin 1986 and references
therein), many correlations now display scatter at the $10-20\%$ level, 
indicating that a high degree of physical uniformity exists even in
this structurally diverse population.
Hydrodynamic simulations of cluster evolution predict tight 
relationships between observable quantities and between those
quantities and the cluster binding mass 
(Evrard 1990; Kang \etal 1994; Navarro, Frenk \& White 1995;
Evrard \etal 1996; Bryan \& Norman 1998), even when some members of
the sample are far from dynamical equilibrium. The existence of both
observed and theoretical correlations implies that the prevalence of
cluster substructure is not a fundamental barrier to interpreting the
properties of the population.

However, moderate biases caused by the presence of substructure are
likely to be present, and we explore the role of substructure in
temperature measurements of the ICM in this paper.  A previous paper
(Mathiesen, Evrard \& Mohr 1999) demonstrated that a ICM clumping
leads to a modest ($\sim 15\%$) overestimate of ICM masses derived
under the typical assumptions of spherical symmetry and isothermality.
As the X--ray data improve, the limits of simplifying assumptions such
as these become clearer.  High-resolution X--ray images reveal
secondary peaks and strong asphericities in many clusters and X--ray
spectra indicate the presence of multiple temperature components
within the cores of many clusters (Fabian \etal 1994, Holzapfel \etal 1997,
Allen \etal 2000).  The {\em Chandra} and {\em XMM} satellite
missions will provide the most detailed maps of the ICM emission
and temperature structure yet obtained and will allow more precise
definition of the limitations of the current models.

Three-dimensional hydrodynamical simulations of cluster formation can
help bridge the gap between the new generation of data and traditional
methods and results. While simulated clusters often do not include
many processes thought to be important to ICM evolution
(e.g. radiative cooling and galactic winds), they excel at the
creation of populations with realistic merger histories (Mohr \etal
1995; Tsai \& Buote 1996).  
Ensembles of simulated clusters can therefore be used to
investigate the effects of accretion events, major mergers, clumping,
and substructure on measurements of the ICM.  In this paper, we
analyze the spectral properties of an ensemble of 24 simulated
clusters using a realistic plasma emission model and assuming a
uniform metallicity $0.3$ times the solar abundance.  We find that
even minor accretion events can significantly bias our measurements of
the mean, mass-weighted cluster temperature, and that clusters
undergoing a major merger can sometimes be identified as extreme
examples of this bias.

Section 2 of this paper describes the simulations, the cluster
ensemble, and the process of creating our spectral images.  Section 3
discusses the various measures of cluster temperature which have seen
frequent use and explores the relationships between them. In
particular we explore the relationship between spectrally determined
temperatures and the mass-weighted mean temperature.  The latter is 
found in simulations to follow most closely the virial relationship.
Section 4 then delves
into cluster dynamics, investigating the effects of a major merger on
the ICM and looking for observable signatures of the merging process.
Finally, section 5 summarizes our conclusions.

\section{Simulations}

We use an ensemble of 24 hydrodynamical cluster simulations, divided
evenly between two reasonable cold dark matter (CDM) cosmological
models.  These models are (i) OCDM ($\Omega_0 = 0.3$, $\sigma_8 =
1.0$, $h = 0.8$, $\Gamma = 0.24$); and (ii) $\Lambda$CDM ($\Omega_0 =
0.3$, $\lambda_0 = 0.7$, $\sigma_8 = 1.0$, $h = 0.8$, $\Gamma =
0.24$).  Here the Hubble constant is $100h$ km s$^{-1}$ Mpc$^{-1}$,
and $\sigma_8$ is the linearly evolved, present power spectrum 
normalization on $8h^{-1}$ Mpc
scales. The initial conditions are Gaussian random fields consistent
with a CDM transfer function with the specified $\Gamma$ 
(e.g. Bond \& Efstathiou 1984). The baryon density is set in each
case to $20\%$ of the total mass density ($\Omega_b =
0.2\Omega_0$).  The simulation scheme is P3MSPH; first a P$^3$M (DM
only) simulation is used to find cluster formation sites in a large
volume, then a hydrodynamic simulation is performed on individual
clusters to resolve their DM halo and ICM structure in detail.  The
baryonic component is modeled with $32^3$ particles, providing a
typical mass resolution of 0.01\% within the virial radius. The
resulting cluster sample covers a little more than a decade in total
mass, ranging from about $10^{14}$ to $3 \times 10^{15}
M_{\odot}$. Further details on these simulations can be found in Mohr
\& Evrard (1997).

The simulations model the dynamical and thermodynamical effects of
gravitation, shock heating and adiabatic work on the ICM.  Several
potentially important pieces of physics are neglected.  Radiative
cooling is one; our clusters cannot produce cooling flows in their
cores. These are inferred to be quite common in the population (Edge
\& Stewart 1991, White \etal 1997), and many methods have been
presented in the literature for removing the effects of cooling flows
from measurements of bulk ICM properties. We expect that the new
satellites {\em Chandra} and {\em XMM} will be able to directly
observe the spatial extent of cooling flows and simply remove those
pixels if desired. We thus analyze our simulations under the
assumption that they are comparable to real ICM observations
correct for the presence of cooling flows through excision (e.g. Markevitch
\etal 1998) or explicit modeling (e.g. Allen \etal 2000) of the
extra emission component.
We also lack any treatment of ICM evolution through galaxy
interactions such as supernova enrichment.  Instead, we assume that 
the abundance of heavy elements is constant throughout a cluster's
history.  The heating effects of galactic winds are difficult to
parameterize accurately, and we postpone this investigation to a future
set of numerical experiments including an approximate version of
galactic feedback. 

It also has been proposed that the ICM electron temperature
lags behind the ion temperature to a significant degree, an effect
which occurs due to the long equipartition timescale for plasmas with
typical ICM densities and temperatures.  This view has been supported
by observation of ASCA temperature profiles (Markevitch \etal 1996),
analytical modeling of accretion shocks (Fox \& Loeb 1997), and
hydrodynamical simulation of the formation of a rich cluster
(Chi\`{e}ze, Alimi \& Teyssier 1998; Takizawa 1999).  
The latter show that the effect is small ($\lesssim 10\%$) 
except in the outer regions of the most massive clusters.  
Our simulations, which treat the ICM as a
single thermodynamic fluid, do not take into account this effect.
Allowing cooler electron temperatures in the outskirts of clusters
would increase the magnitude of the spectral effects calibrated below.

\subsection{Spectral image generation}

From the simulation output, observables such as X--ray spectra
and surface brightness maps are
obtained as follows.  Each simulation particle is given a model for
its X--ray emission based on its local ion density $n_i$ and thermodynamic
temperature $T_i$.  Previous works have traditionally used a simple
Bremsstrahlung-based model such as $\epsilon_i \propto
n_i^2T_i^{1/2}$, but we employ the more sophisticated {\texttt mekal}
model from the XSPEC utility (Mewe \etal 1986; a URL for the XSPEC
user's manual is also in the bibliography).  The {\texttt mekal}
spectra are created using 0.3 solar abundance for elements heavier
than hydrogen. Each cluster is ``viewed'' along a single axis, and a
subvolume of the simulation containing the cluster is divided into
image pixels.  Each particle's emission is collected into the image
and smoothed over neighboring pixels with a Gaussian function of the
same width as that particle's SPH smoothing kernel.  The resultant
``data cube'' contains a spectrum at each ``sky'' position.  To our
knowledge, this is the first such spectral visualization of a
simulated intracluster medium.

The particle data are output at twenty time intervals over the entire
simulation, equally spaced in cosmic time.  Groups are identified at
each epoch using a friends-of-friends algorithm with linking parameter
0.15 the mean interparticle separation.  This procedure connects
particles having separations corresponding to roughly $\gtrsim 400$
times the background density (Lacey \& Cole 1994). The position of
each group is taken to be that of the dark matter particle with the
minimum gravitational potential; this point is always used as the
image center.  We have created a catalogue of ``major merger'' events
for each cluster by identifying all instances of groups with two or
more progenitors having a mass ratio $\ge 10\%$.  The entire sample
contains 75 such events.  Images are created for the last 16 outputs
along a single viewing direction,
resulting in a total of $16 \times 24 = 384$ observable physical
states. In certain specified situations we use a subset of these
outputs, focusing on the images corresponding to redshifts about equal
to 0, 0.5, and 1 for each cluster.  The full evolutionary histories of the
clusters are used to enhance the variation of substructure in our
sample and provide insight into the effects of mergers on the ICM's
observable characteristics.

Our primary database consists of two spectral cubes for each output
differing only in radial scale.  We choose half-widths of $r_{500}$
and $r_{200}$ in the plane of the sky, and line-of-sight depth of
$2r_{200}$.  The scale $r_{\deltac}$ is the radius at which the mean
interior density contrast falls to $\deltac$ times the critical density
$\rho_c \sequiv 3 \Ho^2/8\pi {\rm G}$
at the viewing epoch. The smaller radius $r_{500}$ corresponds roughly
to the readily visible portion of current observations, while
$r_{200}$ is the theoretical preference for the virial radius of
clusters in an $\Omega = 1$ cosmology. We used scaled radii rather
than a fixed metric radius in order to probe
regions with similar dynamical character.  Each
image uses the {\texttt mekal} spectral model for the ICM X--ray
emission, and contains a 256-pixel spatial image of the cluster for
each energy bin. The model spectra have a bin size of 100 eV and 
bandpass of [0.1,20] keV, allowing a great deal of flexibility in
producing spectra comparable to observations.  The spectral cubes
created for this work have a bin size of 150 eV and a bandpass of
[0.5,9.5] keV in order to simulate the quality of data expected
from {\em Chandra's} Advanced CCD Imaging Spectrometer (ACIS).
A color figure (Table 2) displaying X-ray surface brightness
and spectral temperature images in two different bandpasses
is presented and discussed in section 4.1, but the reader may
wish to glance at it now to get a visual impression of the data
used in this paper.

\begin{figure}
\epsfxsize 6.0in \epsfysize 6.0in \epsfbox{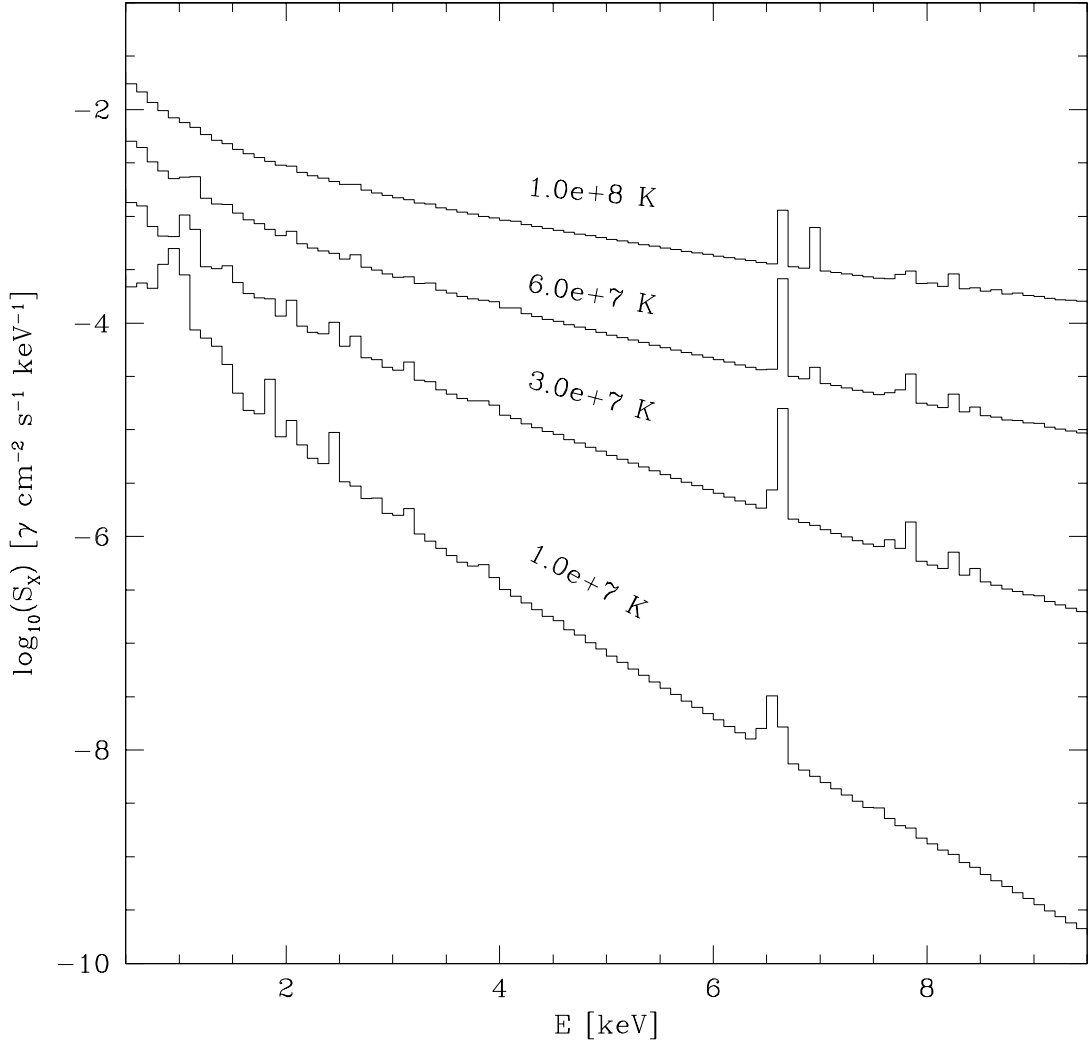}
\caption[Sample {\texttt mekal} spectra at 0.3 solar abundance and
typical ICM temperatures]{Sample {\texttt mekal} spectra at 0.3 solar
abundance.  The $1.0 \times 10^8$~K (8.6 keV) spectrum is plotted in
its true units on the y-axis, while the cooler spectra are scaled
downward by factors of 4 ($6.0 \times 10^7$~K), 16 ($2.0 \times
10^7$~K), and 100 ($1.0 \times 10^7$~K). These temperatures correspond
to $kT =$ 8.6, 5.2, 1.7, and 0.86 keV.  The flux was calculated
assuming a 1 Mpc$^3$ volume of gas with hydrogen density
$10^{-3}$ cm$^{-3}$ at a distance of 225 Mpc.}
\label{spectra}
\end{figure}

As we will see, our use of a spectral model for ICM emission has
important consequences.  Previous X--ray images derived from
simulations use a simple Bremsstrahlung model, but line emission can
make up as much as 20\% of the cluster luminosity for gas at
relatively low ($\lesssim 2.0$ keV) temperatures.  While continuum
emission in the low energy ROSAT passband is a weak function of
temperature above this threshold (Mohr \etal 1999), the introduction
of line emission into the model brings a strong temperature dependence
back into the soft X--rays. This effect can be seen in Figure
\ref{spectra}, which plots isothermal {\texttt mekal} spectra with 0.3
solar abundance at temperatures typical of the ICM.  Since soft ($E <
2$ keV) line emission is progressively more pronounced at cooler
temperatures, even a small amount of cold gas in the observation
window can make a spectrum appear significantly cooler. This effect
is central to the results presented in this paper.

\section{Temperatures}

Cluster atmospheres are only approximately, never identically,
isothermal.  Description of the ICM by a single temperature $T$ is
not only incomplete but also dependent on the measure employed to
derive the value of $T$.  Common definitions employed in the
literature use the mass-weighting and an approximate emission-weighting
methods described below.  

With Lagrangian simulations, volume
integrals tranform to sums over particles within the volume
\begin{equation}
\int dV \, \rho^n \, T \ \longrightarrow \ \sum_i \, m_i \, \rho_i^{n-1} \, T_i
\end{equation}
where $m_i$ is the mass and $\rho_i$ and $T_i$ the density and
temperature of the $i^{\rm th}$ particle.  Our simulations employ 
equal mass gas particles, so the mass-weighted temperature within 
$r_{\deltac}$ is a simple average over $N$ SPH particles 
\begin{equation}
T_m \ = \ \frac{1}{N}\sum_i^N \, T_i, 
\end{equation}
while the approximate emission-weighted temperature employs a $\rho^2$
volume weighting that transforms to 
\begin{equation}
T_e \ = \ \sum_i^N \, \rho_i T_i \, / \, \sum_i^N \, \rho_i
\end{equation}
in a Lagrangian implementation.  Somewhat more sophisticated 
implementations of emission-weighting have been employed;  for
example, Evrard, Metzler \& Navarro (1996) use a ROSAT emission
weighting measure that incorporates the temperature dependence of
bremsstrahlung emission within an $0.1-2.4$ keV energy passband.
Within the simulations, we determine both $T_m$ and $T_e$ within
spherical boundaries of radii $r_{500}$ and $r_{200}$.

If cluster atmospheres within $r_{\deltac}$ are hydrostatic and 
in virial equilibrium, then the thermal energy of the ICM will reflect
the net gravitational potential energy of the cluster.  This leads to
the oft-cited virial relation between mass-weighted temperature and
binding mass 
\begin{equation}
kT_m \ = \ \alpha GM_{\deltac}/r_{\deltac}
\end{equation}
where $M_{\deltac}$ is the total mass within a sphere of radius
$r_{\deltac}$ and $\alpha$ is a dimensionless form factor.  Assuming
that $\alpha$ is a constant independent of mass and redshift (strict
self--similarity) and recalling that $M_{\deltac}(z) \propto
\deltac\rho_c(z)r_{\deltac}^3$ with $\rho_c(z) \propto H^2(z)$, 
we arrive at a compact form for the virial mass-temperature relation
\begin{equation}
h(z)M_{\deltac} \ \propto \ T_m^{3/2}
\end{equation}
with $h(z)=H(z)/100 \kmsmpc$. The factor $H(z)$ takes into account
the cosmological scaling of this relationship, and can be derived
from the Friedmann equation. For the flat universe modeled in our
simulations, 
$(H(z)/H_0)^2 = [\Omega_{\rm m}(1+z)^3 + \Omega_\Lambda]$,
where $\Omega_\Lambda = \Lambda/3H_0^2$. This factor is equivalent
to the evolution factor $E(z)$ used by Bryan \& Norman (1998).

\begin{figure}
\epsfxsize 6.0in \epsfysize 6.0in \epsfbox{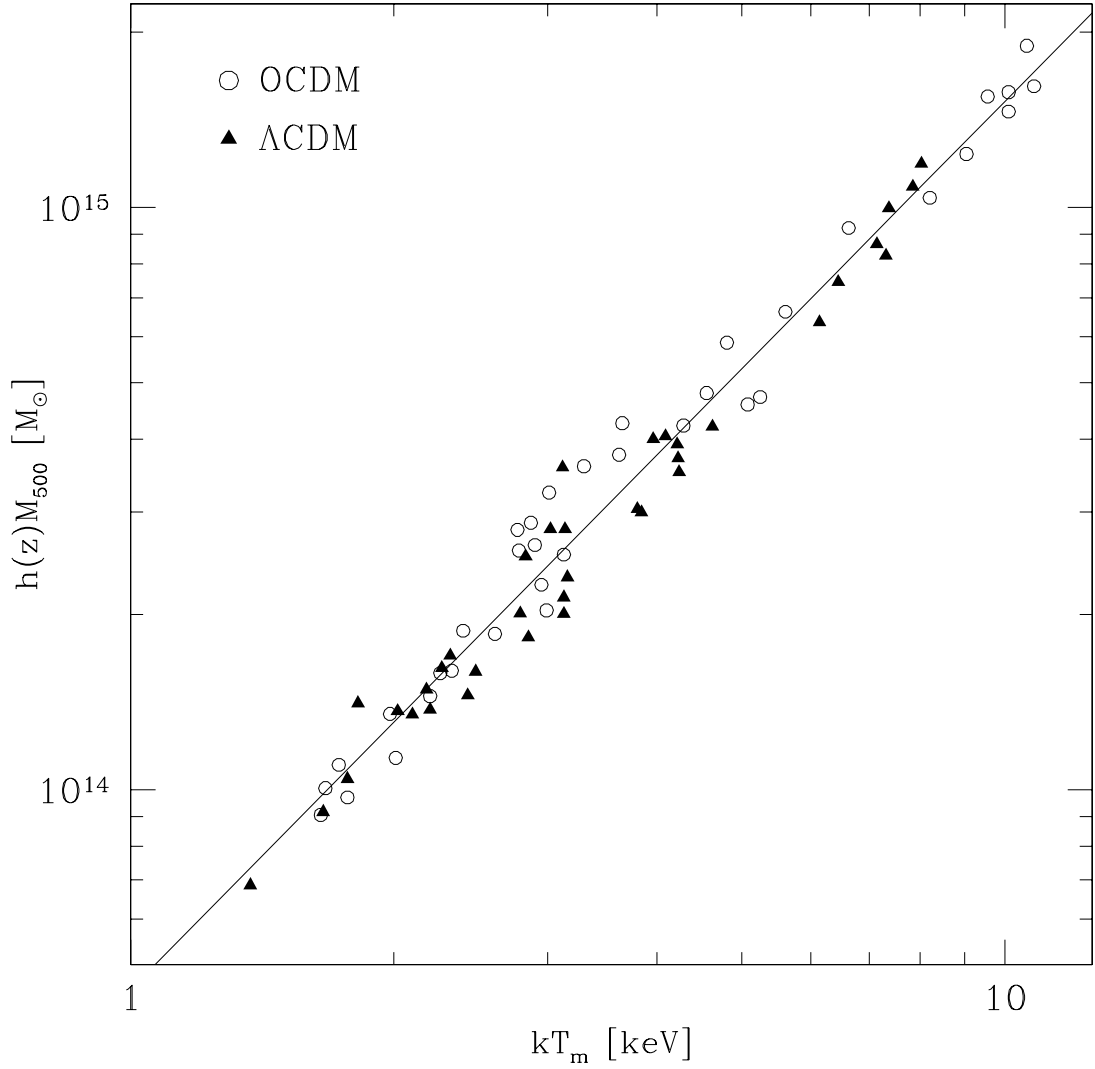}
\caption[The virial mass-temperature relationship within $r_{500}$ for
simulated clusters.]
{The virial mass-temperature relationship within $r_{500}$ for simulated
clusters.  The solid line represents the least-squares fit to the data 
given in equation 6. Three output frames per simulation are included in
the plot, corresponding to the redshifts of 0, 0.5, and 1.0.}
\label{virmasstemp}
\end{figure}

This relationship determined at $\deltac \se 500$ is presented 
in Figure \ref{virmasstemp} using three simulation outputs for
each cluster at the redshifts $z = 1$, $0.5$ and 0.  The data agree
well with the virial expection;  the best-fit power law
\begin{equation}
h(z)M_{500} = (1.2\pm 0.1)\times 10^{15} (T_m/10 \kev)^{1.52\pm 0.03}
[{\rm M}_\odot]
\end{equation}
exhibits only 14\% scatter in mass at fixed $T_m$.
The data within  $\deltac \se 200$ have a similar slope and
slightly larger scatter.

Results such as these have led many cosmologists to use
the virial relation, calibrated by simulations or a
well-studied nearby cluster, to estimate binding masses from cluster
temperatures.  Even if current simulations are complete and exact in
their physical description of the ICM, this procedure need not be
correct, since the mass-weighted temperature is not a directly
obervable quantity.
Computational cosmologists have often employed the simplified 
emission-weighted temperature $T_e$ in an effort to move closer to
observations. What is actually measured, however, is a spectral temperature
derived from  fitting a single Raymond-Smith or {\texttt mekal}
model to the spectral data of a cluster.  Papers analyzing a
single cluster or a small sample sometimes employ two temperature
components and/or a model cooling flow spectrum in order to improve the
fit.  These variations are physically well-motivated but often
unnecessary in the sense that a single temperature component usually 
provides an acceptable fit to even high quality spectral data
(Holzapfel \etal 1997).

\subsection{Spectral Temperatures}

From the simulations, we produce spectral temperatures by fitting 
isothermal models to sample photon spectra.  The latter are produced
by summing emission from gas within a cylinder of 
appropriate radius and depth $2r_{200}$ centered on the cluster. This
collection volume includes most of the line-of-sight gas in the
simulation, but projection effects from large-scale structure
(represented in the high-resolution simulation by massive dark matter
particles with no associated baryons) are not taken into account.
The combined spectrum is rebinned into channels of width 150 eV
(comparable to the {\em Chandra} ACIS energy resolution), convolved
with the {\em Chandra} effective area function, and assigned Poisson
error bars appropriate to the number of photons in each bin, assuming a
total photon count of 20,000.  Bins with a low count rate are grouped
together until they contain at least twenty photons. The result is
then fit to a single-temperature {\texttt mekal} spectrum with 0.3 solar
metallicity. No Poisson
{\em noise} is added to the photon counts, as we are trying to
isolate the systematic effects of substructure, but the error bars are
used in fitting to assign appropriate relative weights to the data. 
The resulting spectral temperature $T_s$ which minimizes the
chi-squared (the normalization of the spectrum is also a
free parameter, of course)
is a direct analog of observed temperatures obtained by the {\em Chandra}
observatory.  Other satellites can readily be simulated by
changing the bin width, bandpass, and effective area function.

\begin{figure}
\epsfxsize 6.0in \epsfysize 6.0in \epsfbox{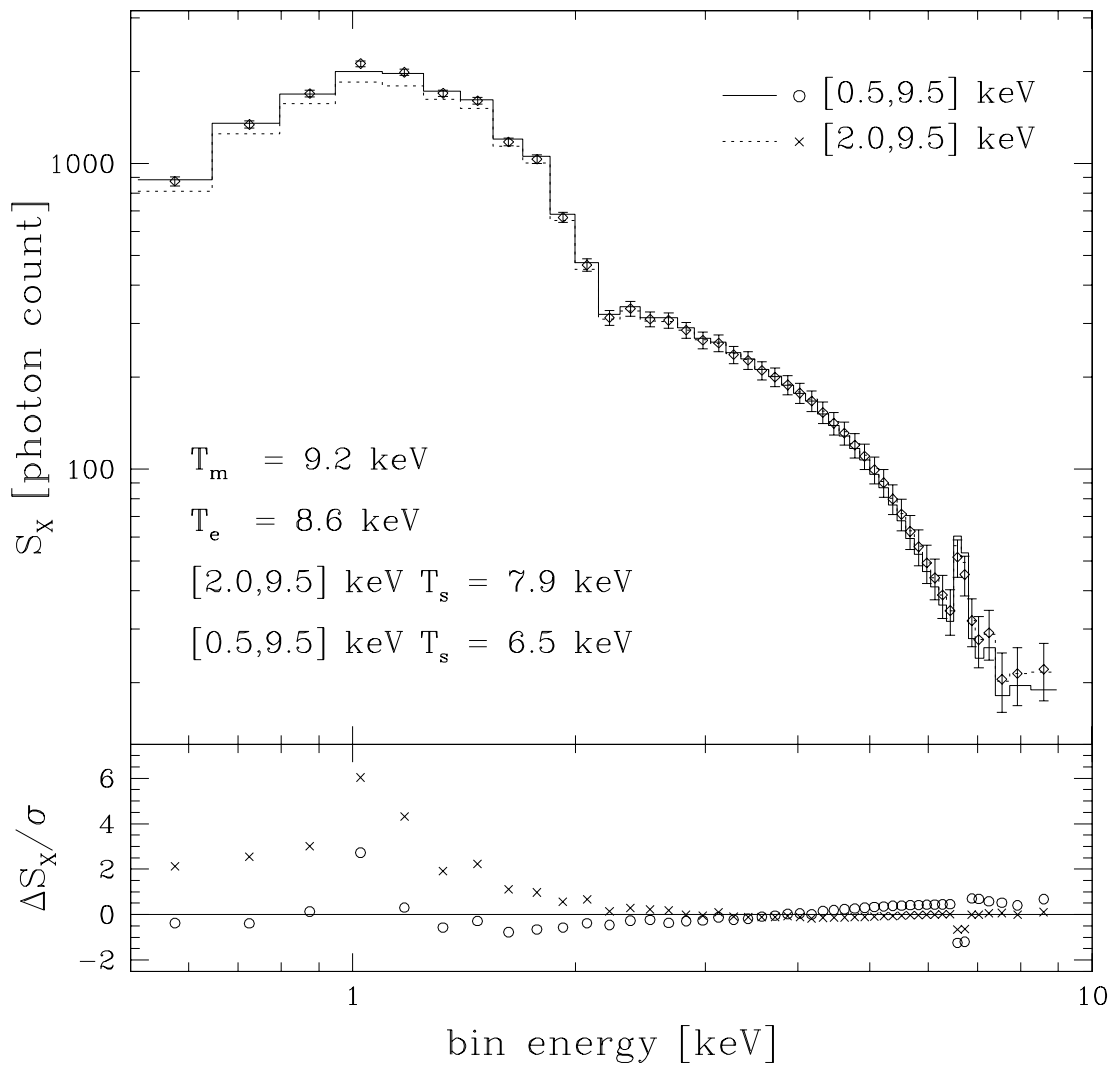}
\caption[Sample fit to the combined spectrum of a simulated cluster.]
{Simulated {\em Chandra} spectrum for a rich
cluster at redshift of 0.042.  The photons are collected
within a cylinder of diameter and depth $r_{200}$ centered on the
minimum potential of the simulation.  The exposure time is scaled
so that the spectrum contains approximately 20,000 photons, but
the best-fit temperatures obtained are insensitive to the total
photon count. Two isothermal mekal spectra are displayed: the solid
line is the best fit for all the bins, and the dotted line is the best
fit that results for bins over 2 keV. The reduced chi-squared parameters
for the two models are $\chi^2/46=0.38$ for the full spectrum and
$\chi^2/36 = 0.05$ for the fit to bins above 2 keV.}
\label{specfit}
\end{figure}

A sample of this process is shown in Figure \ref{specfit}, which
displays the combined spectrum, best-fit spectrum, and residuals for a
typical cluster of the ensemble. Although this spectrum is formally
well fit by a single-temperature model, there are clear trends in the
residuals: the strength of the iron line is somewhat too low for a
continuum with this slope, the width of the iron line is increased
somewhat from the lower-energy transitions in this complex, the 1.0
keV bin is much too high, and the overall shape of the spectrum is
slightly different. With the exception of the iron line and the 1.0
keV bin these excursions are very small, and we would not expect them
to be visible in a real spectrum with this photon count and typical
noise levels. Deep exposures from {\em Chandra} might be able to pick
them out, however.  Increasing the photon count to 100,000 for this
cluster brings the reduced chi-squared up to 1.64 for 54 degrees of
freedom, enough to detect the difference in spectral shape even with
typical levels of random error.  The unusual coincidence that a
realistic combined spectrum has nearly the same shape as an isothermal
spectrum may be explained by our mediocre energy resolution, which
keeps us from constraining the structure of the emission lines to a
useful degree.

\begin{figure}
\epsfxsize 6.0in
\epsfysize 6.0in
\epsfbox{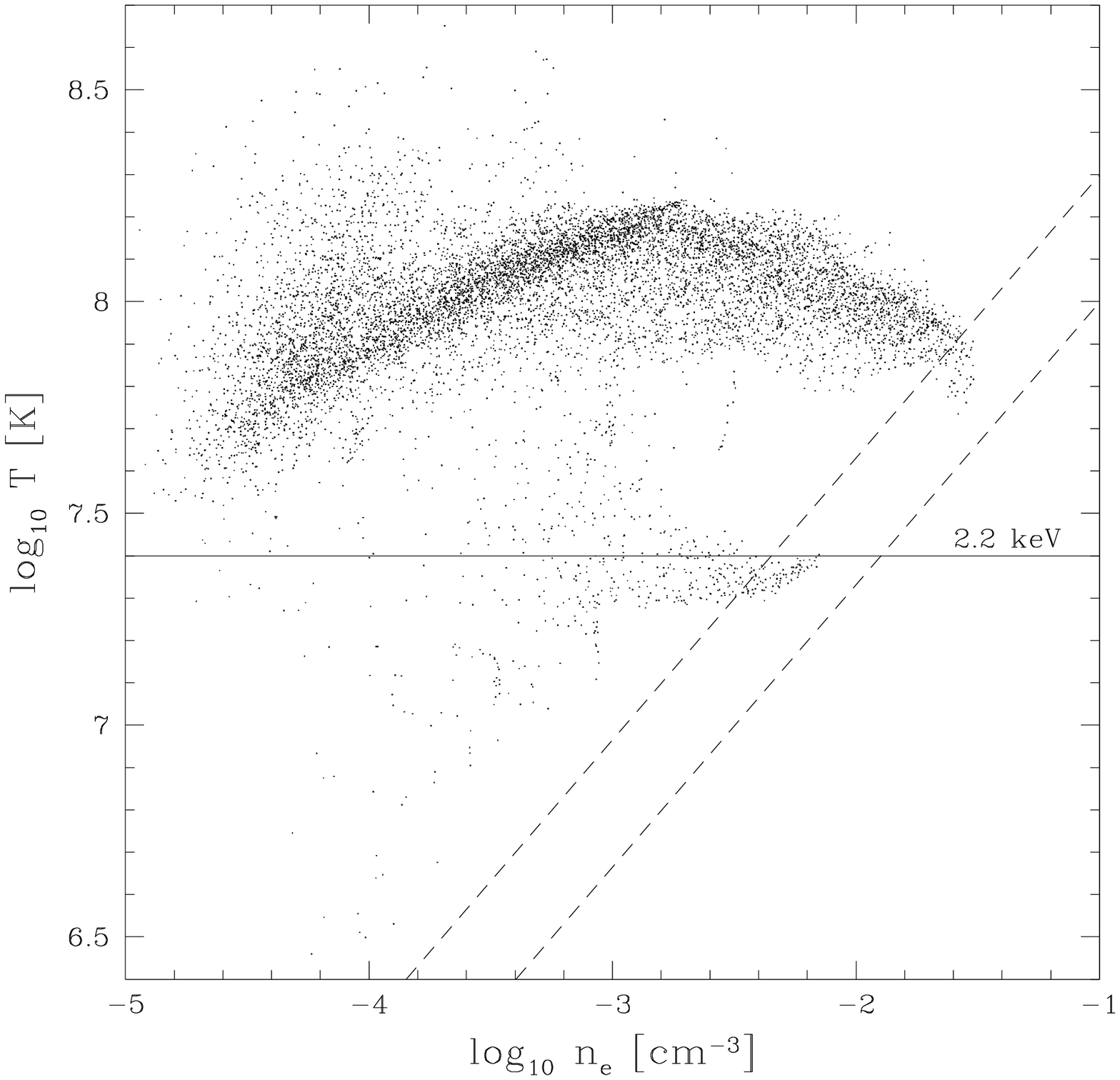}
\caption[Phase-space diagram for the cluster of Figure \ref{specfit}]
{Phase-space diagram showing the local temperature and number density of
particles lying within $r_{200}$ of the cluster whose spectrum
is displayed in Figure \ref{specfit}. 
The horizontal line is drawn to separate out the merging subclump
and indicate an approximate threshold for significant line
emission in the [0.5,2.0] keV band. Particles below this line
make up only 4\% of the total mass, but cause a 30\% shift in
the spectral temperature. The dashed lines represent 
minimum entropy thresholds of 40 and 80 keV cm$^2$ (Lloyd-Davies
\etal 2000);  if the ICM were preheated prior to cluster formation
infalling subclumps would be tend to be more diffuse.}
\label{ob1phase}
\end{figure}

We have cut off the spectrum at 9.5 keV due to a lack of photons above
this range.  The spectral temperatures are significantly
lower than both the emission- and mass-weighted temperatures for this
cluster.   The reasons for these differences are 
related to this particular cluster's dynamical state.  In 
Figure \ref{ob1phase}, we show a phase diagram of the ICM gas within 
$r_{200}$ the cluster center.  The tongue of low temperature material
is a small merging subclump that is barely visible
in the X--ray surface brightness.  We have drawn a
horizontal line at 2.2 keV (log $T$ [K] = 7.4)
to visually separate the core of this subclump.  Particles below this
threshold (and especially those in the clump, which is nearly
as dense as the cluster core) produce significant line emission
in the [0.5,2.0] keV band and enhance the spectrum in this region.
Only 4\% of the particles in the cluster lie below this line, but
it turns out that this is more than enough to significantly alter the 
shape of the spectrum.  The dashed lines in Figure \ref{ob1phase}
represent the minimum entropy thresholds inferred by Lloyd-Davies,
Ponman, \& Cannon (2000) from X--ray observations of galaxy groups,
and will be discussed further at the end of section 3.2.  

This is an important point: minor mergers such as this are just as likely
as major events to cause deviations between the spectral and mass-weighted
temperatures. Not only are they much more common, their gas is colder---and
therefore more luminous in the soft X--ray band.  

\begin{figure}
\epsfxsize 6.0in
\epsfysize 6.0in
\epsfbox{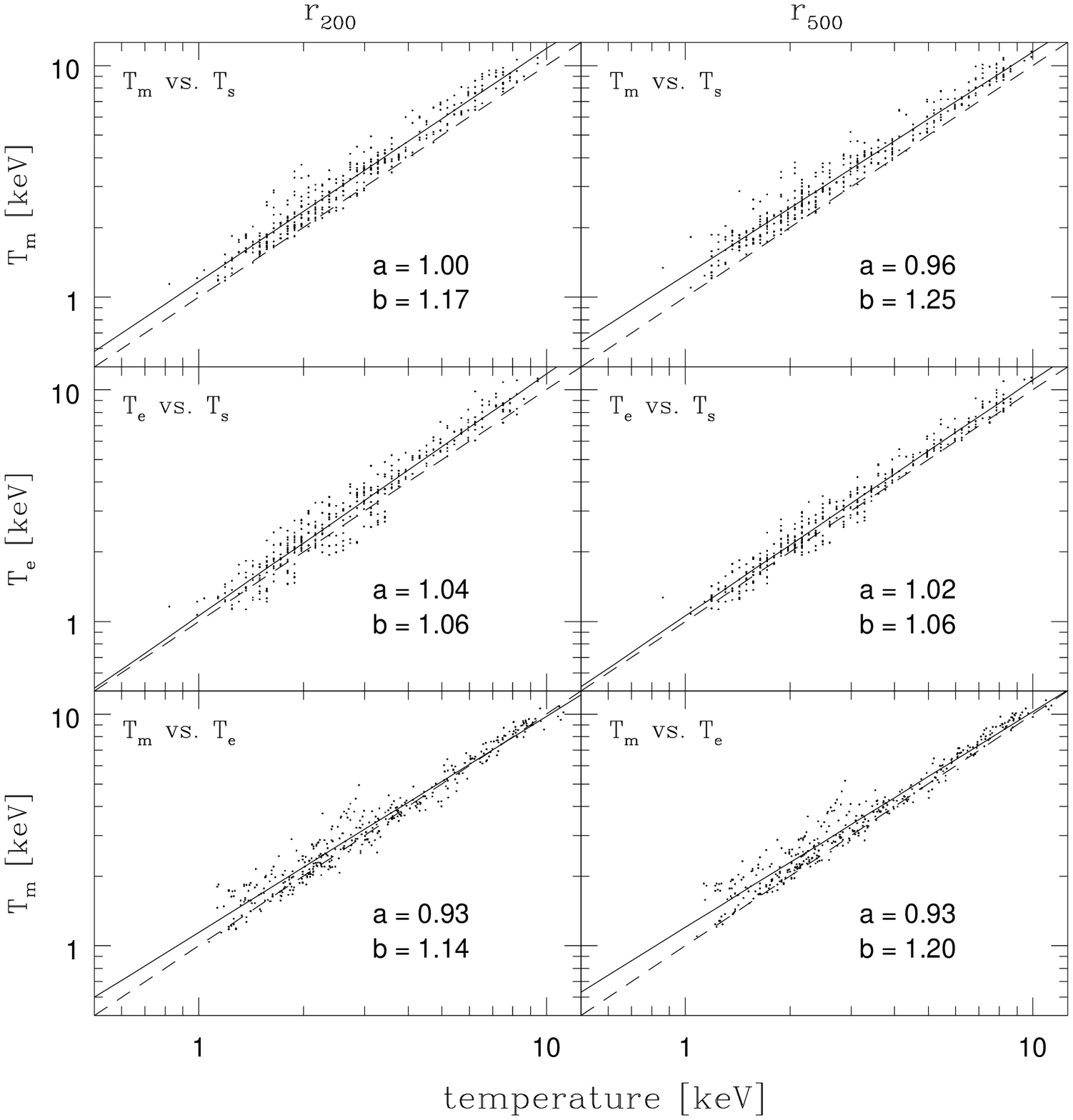}
\caption[Correlations between spectral, mass-weighted, and
emission-weighted temperatures within $r_{500}$ and $r_{200}$.]
{Correlations between spectral, mass-weighted, and
emission-weighted temperatures for 384 simulation outputs
(24 different clusters). Spectral temperatures have been
fit in the 0.5-9.5 keV band. The left column shows measurements
within the virial radius $r_{200}$ while the right column uses
measurements within $r_{500}$. The coefficients $a$ and $b$ relate
the x and y coordinates as $T_y$ = $bT_x^a$. The one-sigma 
uncertainties in $a$ and $b$ are typically 0.010 and 0.015, respectively.
The dashed curve is the line of equality, while the solid
line is the linear, least-squares best fit to the data.}
\label{tvst}
\end{figure}

We find that this bias
is nearly universal in our sample, and are motivated to construct the
mutual correlations of the three temperatures.  These relationships
are displayed in Figure \ref{tvst}. The spectral temperatures occur
at discrete intervals because we tabulated {\texttt mekal} outputs
at temperatures with spacing $\Delta\log T = 0.02$ in all our analysis
(at considerable savings in CPU time). This interval is much
smaller than the uncertainties associated with our spectral
temperatures. The constants $a$ and $b$ displayed in the plots
result from fits $T_y = bT_x^a$ obtained by least-squares
linear regression using logarithmic variables. 

We can see from these plots that the mass- or emission-weighted
temperatures are almost universally higher than the spectral
temperatures.  Why is this the case?  The mass-weighted temperatures,
as we have seen, follow the virial expectation quite closely
and can be used as a benchmark for the cluster's evolution.
The emission-weighted temperatures strongly favor the core
region, so also tend to mark the potential well depth
rather well. The spectral temperature, on the other hand,
is weighted by the fitting process towards whichever regions
emit the most photons. This favors both high-density regions
and cool regions, and the cores of accreting subclumps satisfy
both criteria. Accreting clumps of cool gas can therefore create
an excess of soft photons for some time during and after a merger
event. This process will be discussed in more detail in section 4, but
for now it suffices to realize that accretion and substructure
can have a rather large effect on spectral temperatures. 
$T_s$ doesn't change a great deal when we go from $r_{200}$ to
$r_{500}$, but the mass-weighted temperature increases; this
is why we see a larger bias within $r_{500}$ despite the
fact that the former window includes a larger volume of cool
gas.

Is there evidence of this phenomenon in the literature? 
Accurate temperature
determinations for clusters are available from the satellites
{\em Einstein}, {\em EXOSAT}, {\em Ginga}, and {\em ASCA}, each of
which uses a different bandpass in fitting their X--ray spectra. We
can test for the presence of the effect described here by looking
for a systematic difference between spectral temperatures in 
different energy bands. All of these satellites have effective
upper limits comparable to the spectrum shown in Figure \ref{specfit}
thanks to the relative scarcity of ICM photons above 10 keV. The
{\em Einstein} MPC temperatures of David \etal (1993) are available
for many clusters and use a consistent lower limit of 2 keV, so we use
this data set as our basis. Each of the other three satellites makes
some use of photons below this threshold, and should therefore
return cooler temperatures than the MPC. The {\em EXOSAT} temperatures
(Edge \& Stewart 1991) used here are derived from Raymond-Smith
spectral fits in the range 1.5-9.0 keV, with an additional constraint
coming from unresolved soft photon counts in the approximate range
0.6-1.0 keV. {\em Ginga} spectra (Hayakude 1989) have an energy range
of 1-20 keV for cluster observations (Arnaud \etal 1991) and
typically obtain a single significant bin below 2 keV. 

Some comparisons between the results of these satellites have
already been made in the literature. David \etal (1993) found
reasonable agreement between MPC, {\em EXOSAT}, and {\em Ginga}
temperatures, and note that {\em EXOSAT} temperatures are
typically 0.7 keV less than their MPC temperatures for clusters
between 5 and 8 keV. They also find reasonable agreement between
{\em Ginga} and MPC temperatures, but fail to note that out of 8
clusters only one of the {\em Ginga} temperatures is higher
than the MPC determination. More telling evidence comes from an
analysis of {\em ASCA} data. Markevitch \etal (1998) state
that {\em ASCA} temperatures determined in the [0.5,11] keV
bandpass  were typically lower by about 0.5 keV than
temperatures determined in other ranges. Markevitch \etal
use the [1.5,2]+[2.5,11] bandpass for their published temperatures
in an effort to avoid this systematic (the hole in the spectrum
avoids a region with a difficult point spread function), and
found good agreement between their temperatures and those of other
satellites.

\begin{figure}
\epsfxsize 6.0in
\epsfysize 6.0in
\epsfbox{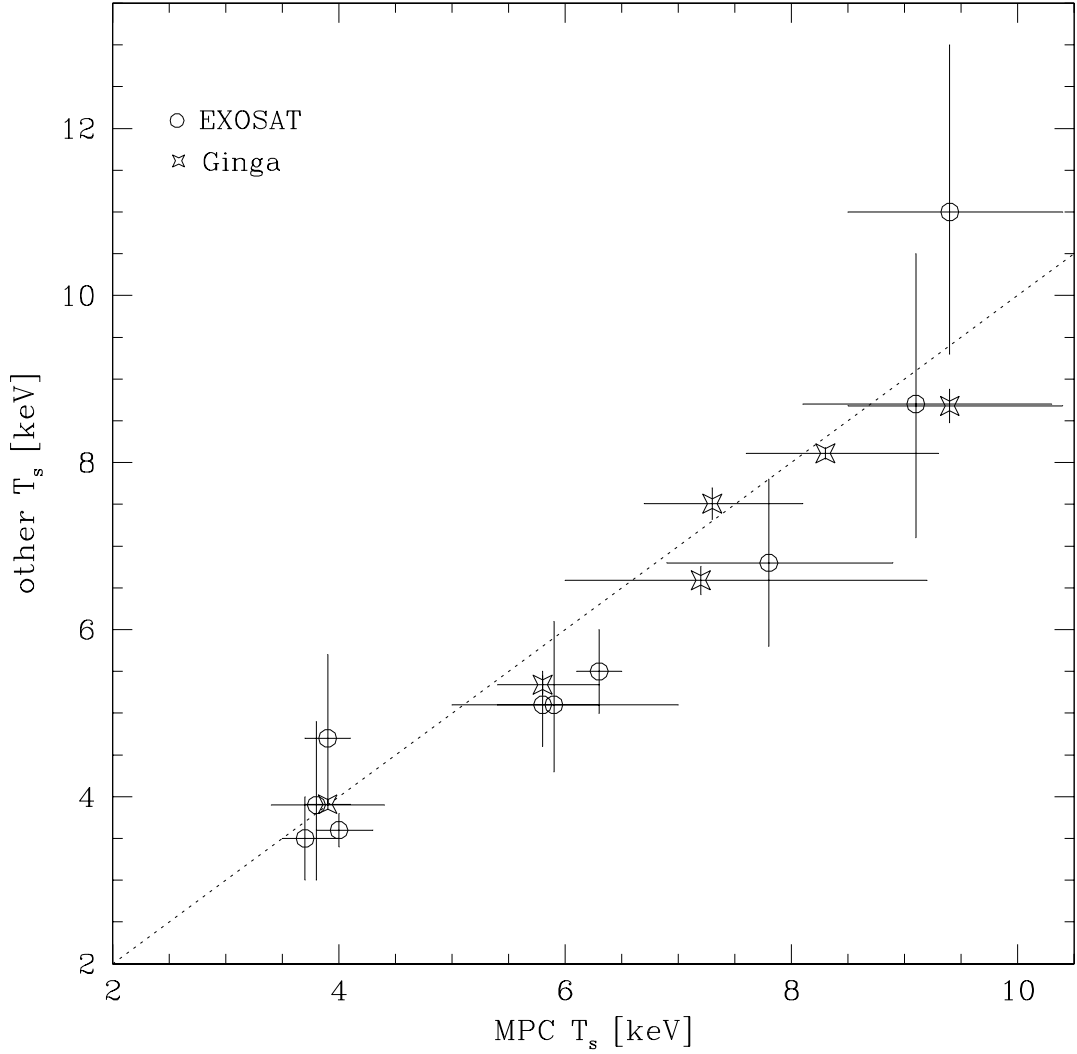}
\caption[Comparison of previously published cluster temperatures.]
{Comparison of previously published cluster temperatures from the
{\em Einstein} MPC, {\em EXOSAT}, and {\em ASCA}. The MPC
temperatures are best-fit spectral temperatures in the [2,10] keV
band, and the other two satellites each make some use of photons
below 2 keV. All error bars represent 90\% confidence intervals.
Clusters with poorly determined MPC ($\delta T > 2.0$ keV) temperatures
are left out because their error ranges easily overlap
the line of equality and their inclusion makes the plot
difficult to read. The probability that only 4 of the 16 data
points scatter above the line is only 4\%, assuming the
temperature determinations are equivalent.}
\label{sattemps}
\end{figure}

If an excess of soft photons is really making the ICM appear cooler
than it is, we should expect that the MPC has the hottest
spectra because it has the highest minimum energy. We thus collect
temperatures from the literature
({\em EXOSAT}: Edge \& Stewart 1991; {\em Ginga}:
Hatsukade 1989 via David \etal 1993; MPC: David \etal 1993;
{\em ASCA}: Markevitch \etal 1998)
and make our own comparison to see if this is true.
There is good overall agreement between {\em ASCA} and {\em MPC}
temperatures, but the gap in the {\em ASCA} spectrum makes
comparison of the techniques difficult; we therefore do not
include these data in our analysis. The {\em EXOSAT} and {\em Ginga}
temperatures used here, however, each make use of the full [2,10] keV
band as well as including constraints from photons below 2 keV.
These temperatures are plotted in Figure \ref{sattemps}, and are for
the most part {\em lower} than the MPC measurements. While in most
cases the error bars intersect the line of equality, the probability
that so many of these observations scatter below the line is quite
small and argues for some systematic error. Assuming that the clusters
are drawn from an unbiased distribution in which each object has equal
probability to fall above or below the line of equality, this uneven
arrangement of clusters (4 above, 12 below) has only a 4\% chance of
occuring at random. It is also interesting to note that all the clusters
whose vertical error bars fall short of the line are in the ``cool''
direction predicted by these simulations.

\begin{figure}
\epsfxsize 6.0in
\epsfysize 6.0in
\epsfbox{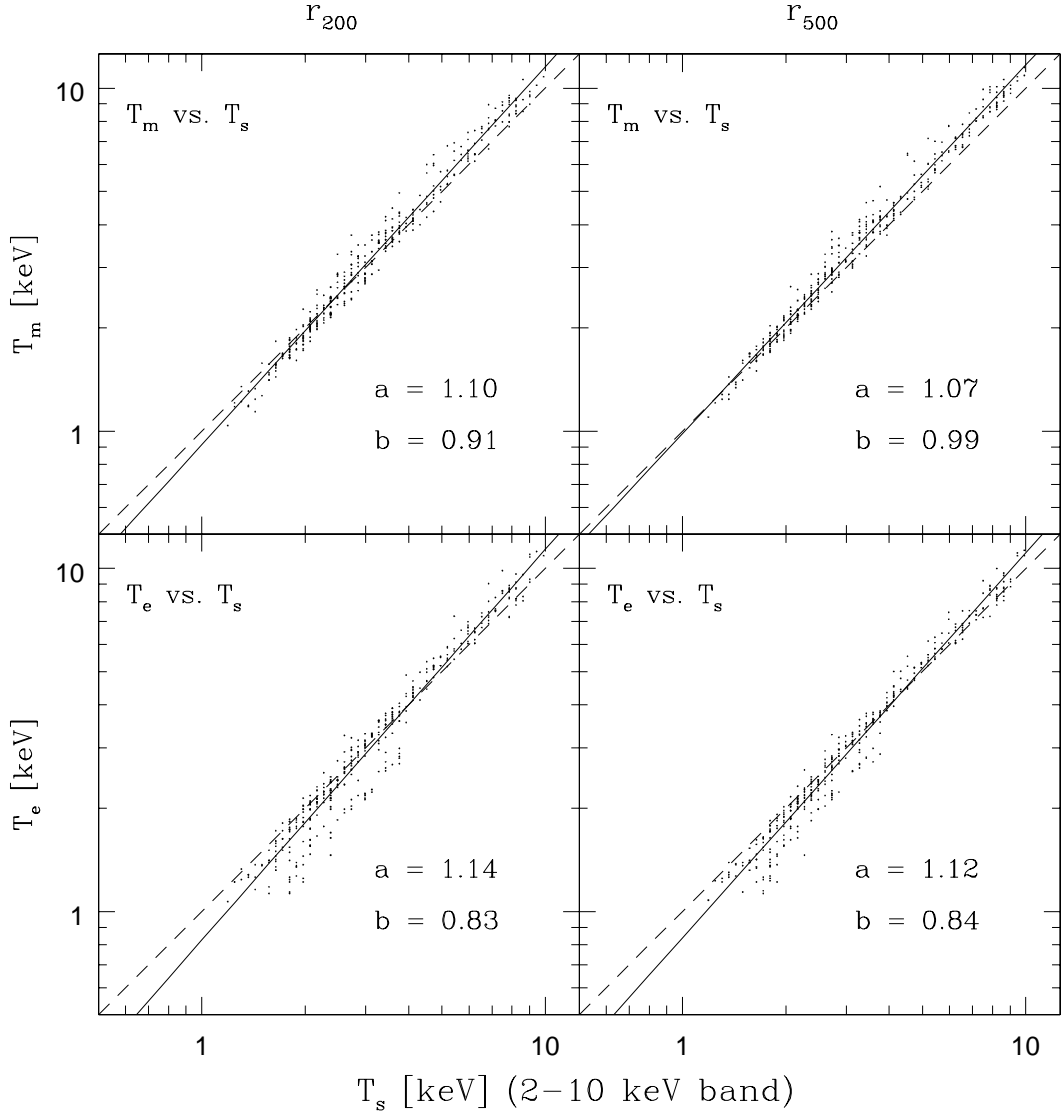}
\caption[Correlations between spectral, mass-weighted, and
emission-weighted temperatures within $r_{500}$ and $r_{200}$;
spectral fitting is in the 2.0-9.5 keV bandpass.]
{Correlations between spectral, mass-weighted, and emission-weighted
temperatures for the cluster outputs. These plots can be read in the
same way as those in Figure \ref{tvst}, except here spectral temperatures
are calculated by fitting the spectrum in the [2.0,9.5] keV bandpass.}
\label{tvst_g2kev}
\end{figure}

The observational evidence, though far from conclusive,
suggests that the influence of line
emission in the [0.5,2] keV range is not negligible.  If true, 
a conservative response would be to adopt the
[2-10] keV band as a standard for spectral temperature
determinations.  Judging from Figures \ref{spectra} and
\ref{specfit}, this would remove most of the excess soft photons
from the spectrum and reduce the bias due to cool gas. We test
the effectiveness of this technique by fitting
our cluster spectra in the [2.0,9.5] bandpass while maintaining
a 20,000 photon normalization over the entire energy range. 
The results of this analysis are shown in Figure \ref{tvst_g2kev};
throwing out all the photons below 2 keV results in temperatures
driven primarily by the Bremsstrahlung component, and the spectral
temperatures are now much closer to the mass-weighted measure.

\begin{figure}
\epsfxsize 6.0in
\epsfysize 6.0in
\epsfbox{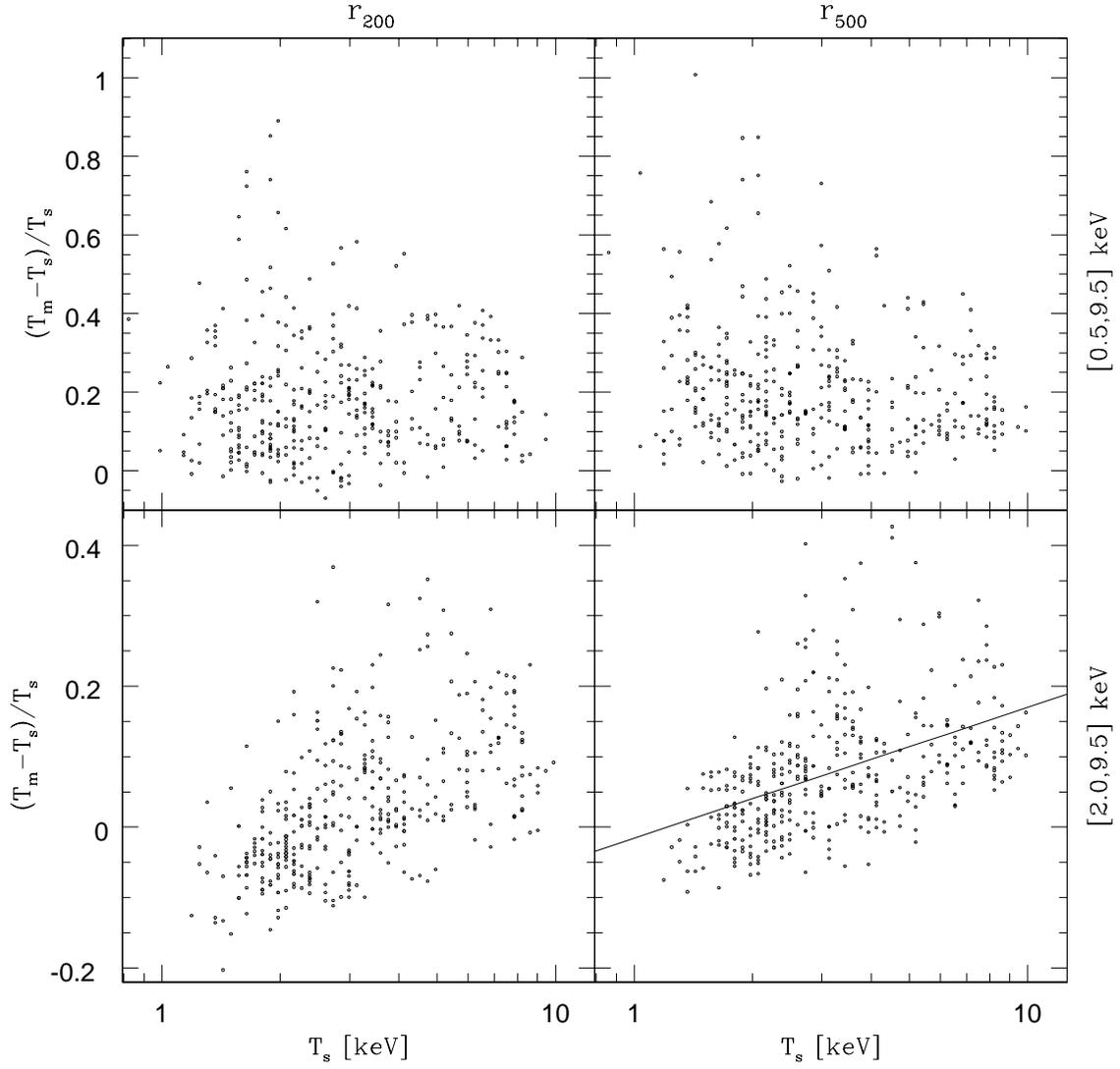}
\caption[The difference between spectral and mass-weighted
temperatures as a function of spectral temperature for
four different methods.]{The fractional difference between
spectral and mass-weighted temperatures as a function of $T_s$
for four different methods.  The upper plots measure $T_s$
within the [0.5,9.5] bandpass, while the lower plots
measure $T_s$ within the [2.0,9.5] bandpass.}
\label{dtvsts}
\end{figure}

Another view of this analysis is given in Figure \ref{dtvsts},
which shows the fractional differenct $(T_m-T_s)/T_s$ plotted
against the spectral temperature for four methods. This plot
makes it easier to estimate the size of the errors and their
scale-dependence. We see that the [0.5,9.5] keV spectral temperatures
almost universally underestimate the mass-weighted temperatures,
typically by 10-20\% but sometimes by as much as a factor of
two.  We see similar results within $r_{500}$ and $r_{200}$
for these spectra.  The picture looks brighter in the [2.0,9.5]
keV bandpass, with almost all the errors confined within $\pm 20\%$,
but the scale-dependence of the effect is stronger.  The panel
which corresponds most closely to real measurements is 
the [2.0,9.5] bandpass temperature for emission within $r_{500}$,
in the lower right of the Figure. The best-fit line drawn through
these points follows the equation
\begin{equation}
\delta T_s \equiv \frac{T_m-T_s}{T_s}=
(0.19\pm 0.02)\log_{10}T_s [{\rm keV}]-(0.016\pm 0.010).
\end{equation}
The uncertainties are at the one-sigma confidence level.
We thus find that even in this bandpass the temperatures of hotter
clusters are underestimated by 10-20\%.  Although gross deviations
are rarer, there are still a few cases where the mass-weighted
temperature is 40\% higher than the spectral temperature.  From
here on, when we use the term ``spectral temperature'' we refer to
the temperature measured within $r_{500}$ in the [2.0,9.5] keV
bandpass, unless otherwise noted.

Before going on, we point out a few effects which are
potentially important but beyond the scope of this work.
First, it seems a likely possibility that electron temperatures
are much lower than ion temperatures in the cluster outskirts. 
Chi\`{e}ze \etal (1998), who performed an Eulerian simulation
of the ICM for a Coma-like cluster, found little difference in the
two temperatures within $r_{500}$ but about a 20\% difference
between the two at $r_{200}$.  Takizawa (1999) performed head-on
collisions of merging clusters designed to produce a final state of 8
keV and found similar results.  Because the Coulomb relaxation time
scales as $T^{3/2}$, any nonequlibrium thermodynamic signal would
affect the scale-dependence of the temperature errors described in Figure
\ref{dtvsts}.  However, the magnitude of this effect is likely to be
small, especially within the higher density $r_{500}$ window.

A second effect which may be important in some clusters
is the absorption of soft X--rays by galactic hydrogen; if
this absorption factor is strong enough the influence of
bins below 2 keV on the fit can be greatly reduced.
Applying a photon absorption model appropriate
for a column density of $3.4 \times 10^{20} {\rm cm}^{-2}$ to our
spectra and rederiving Figures 5, 7, and 8, we find no significant
change in the best-fit parameters for these correlations.
However, a column density of 
$10^{21} {\rm cm}^{-2}$ is sufficient to
change the behavior of the spectral temperatures.
The best-fit parameters for Figure \ref{tvst} under this level
of absorption become $T_m = 1.06T_s^{1.03}$ and $T_e = 0.96T_s^{1.06}$
within $r_{200}$; and $T_m = 1.12T_s^{1.00}$ and $T_e = 0.95T_s^{1.06}$
within $r_{500}$. The results for the [2.0,9.5] keV band spectral
temperatures are unchanged under both absorption models.

We close this section with a brief discussion of preheating.
The slope of the $L_X$--$T$ relation in cooling-flow corrected clusters
is well-constrained observationally to be 2.64 +/- 0.27 
(Markevitch 1998),
a value significantly different from the slope of 2 predicted
by models of self-similar gravitational collapse. Simulations
which preheat the baryons at an early epoch impose a minimum
entropy on the ICM during cluster formation and have been able
to better reproduce the observed slope (Cavaliere \etal 1998;
Bialek, Evrard \& Sulkanen in preparation).  If preheating proves
to be important, 
the density (and therefore emissivity) of very cool infalling
gas will be limited and its influence on the spectrum reduced.
Two such thresholds are plotted in Figure
\ref{ob1phase} to illustrate this effect. These lines represent
minimum entropies (defined as $T/n_e^{2/3}$) of 60 and 120 keV cm$^2$,
reported in a paper by Lloyd-Davies, Ponman, \& Cannon (2000) which
directly measures the entropy in 20 cluster cores, including 4 groups
with spectral temperatures below 1 keV. Because the baryon fraction
in these simulations is about a factor of two higher than
observed ICM mass fractions, we rescale these entropy thresholds
by a factor of $2^{-2/3}$ to 40 and 80 keV cm$^2$.

The effect of preheating on our results
is difficult to assess without recourse to another simulation
ensemble, but we can estimate the magnitude of this effect by
assigning each particle violating the upper threshold
the maximum allowed density at its temperature. Applying this
technique to the cluster displayed in Figure \ref{ob1phase}
results in a spectrum with best-fit temperatures
$kT_s = 7.5{\rm keV}$ (an increase of 15\%) in the [0.5,9.5] keV band and
$kT_s = 8.2{\rm keV}$ (an increase of 4\%) in the [2.0,9.5] keV band.
We can conclude that a large degree of preheating has
the potential to reduce the deviation between spectral and
mass-weighted temperatures, especially in the [0.5,9.5] keV band.
We defer further analysis of this question in
anticipation of a preheated simulation ensemble
calibrated to reproduce the observed scaling relations
(Evrard, Mohr \etal, in preparation).

\subsection{Observable Consequences}

\begin{figure}
\epsfxsize 6.0in
\epsfysize 6.0in
\epsfbox{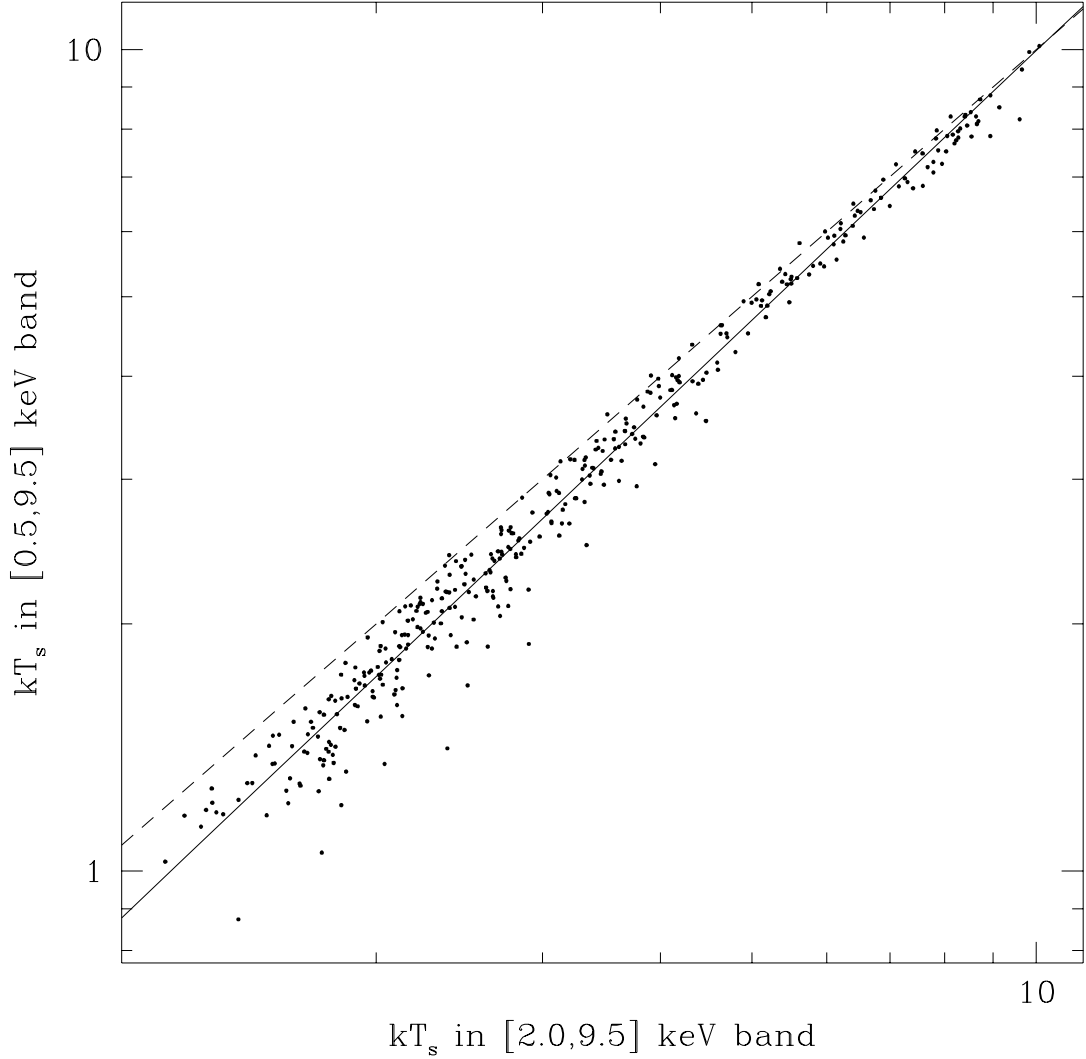}
\caption{The relationship between spectral temperatures 
fitted in the [0.5,9.5] keV and [2.0,9.5] keV bandpasses.
This relationship should be reproducible with Chandra
ICM observations.  The solid curve is the best-fit line
described in the text, and the dashed is the line of
equality.  The mean scatter around the best-fit relation is
0.033 in $\log(kT)$, or 8\%. The temperatures
are calculated within $r_{500}$. The best-fit relation
within $r_{200}$ has a slightly smaller slope (1.06 rather
than 1.09) but the same normalization and scatter.}
\label{ts_tsg2}
\end{figure}

The results of the previous section make clear two important points.
First, spectral temperature determinations tend to be
lower than the underlying mass-weighted temperature. Because of
this bias, an analysis of the population which makes use of
spectral temperatures and the virial relationship will underestimate
the binding masses of rich clusters by 15-30\%.
Second, the exact nature of the systematic is dependent
on the bandpass used in spectral fitting.
A simple test of this dependence can be easily made with
Chandra data by comparing [0.5,9.5] keV band temperatures to
[2.0,9.5] keV band temperatures.  Predictions for this effect from 
the simulation ensemble are shown in Figure \ref{ts_tsg2}.
The solid line in this figure is the linear least-squares
best fit to the points, and follows the relation
\begin{equation}
kT_{s, {\rm [0.5-9.5 keV]}} = (0.81 \pm 0.01)\ kT_{s, {\rm [2.0-9.5 keV]}}
^{1.09\pm0.01} .
\end{equation}

An interesting side effect of this spectral modeling becomes
apparent when we create azimuthally averaged surface
brightness profiles for our simulations in the [0.5,2.0]
keV bandpass. Analysis of {\em ROSAT} position sensitive
proportional counter (PSPC) images has shown that cluster emission
profiles are generally well fit by a beta-model with an average slope
parameter $\beta$ of 0.64 (Mohr \etal 1999, this paper is referred
to as MME in this section);
this value is consistent with previous analyses and the canonical
value $\beta = 2/3$. The value of this parameter has been a challenge
for simulations, which have traditionally produced much steeper
emission profiles.  An earlier generation of surface brightness
images created from this ensemble, for example,
produced profiles which were well fit by a beta-model but with
$\langle\beta\rangle = 0.83$; such results are typical in the
field (e.g. Eke \etal 1998), but not often discussed. 

Simulated clusters have moderate temperature profiles;
the (mass-weighted) ICM temperature at $r_{500}$ is typically
between 80 and 90\% of the mean temperature within this radius.
The gas in the outer regions also tends to exhibit wider
variations in density and temperature (Mathiesen \etal 1999),
and pockets of gas much cooler than this can be found in most
clusters.  The luminosity of simulated clusters is generally
modeled with a bremsstrahlung spectrum, which is rather
insensitive to temperature in a soft X--ray band such as the 
[0.5,2.0] keV used to analyze PSPC images in Mohr \etal
When a spectral model including line emission is used instead,
we find that cool gas in the outskirts of the cluster (including
the very cool gas near the virial radius which lies along
our line-of-sight) significantly enhances the luminosity in
this band. The mean $\beta$-parameter for our ensemble's 
spectral profiles is 0.59, and corresponds much more closely
(in both value and process) to observed clusters. This sudden
reversal of a long-standing discrepancy between simulations and
reality emphasizes the importance of interpreting simulated data
with a realistic ``observational'' framework.

\begin{figure}
\epsfxsize 6.0in
\epsfysize 6.0in
\epsfbox{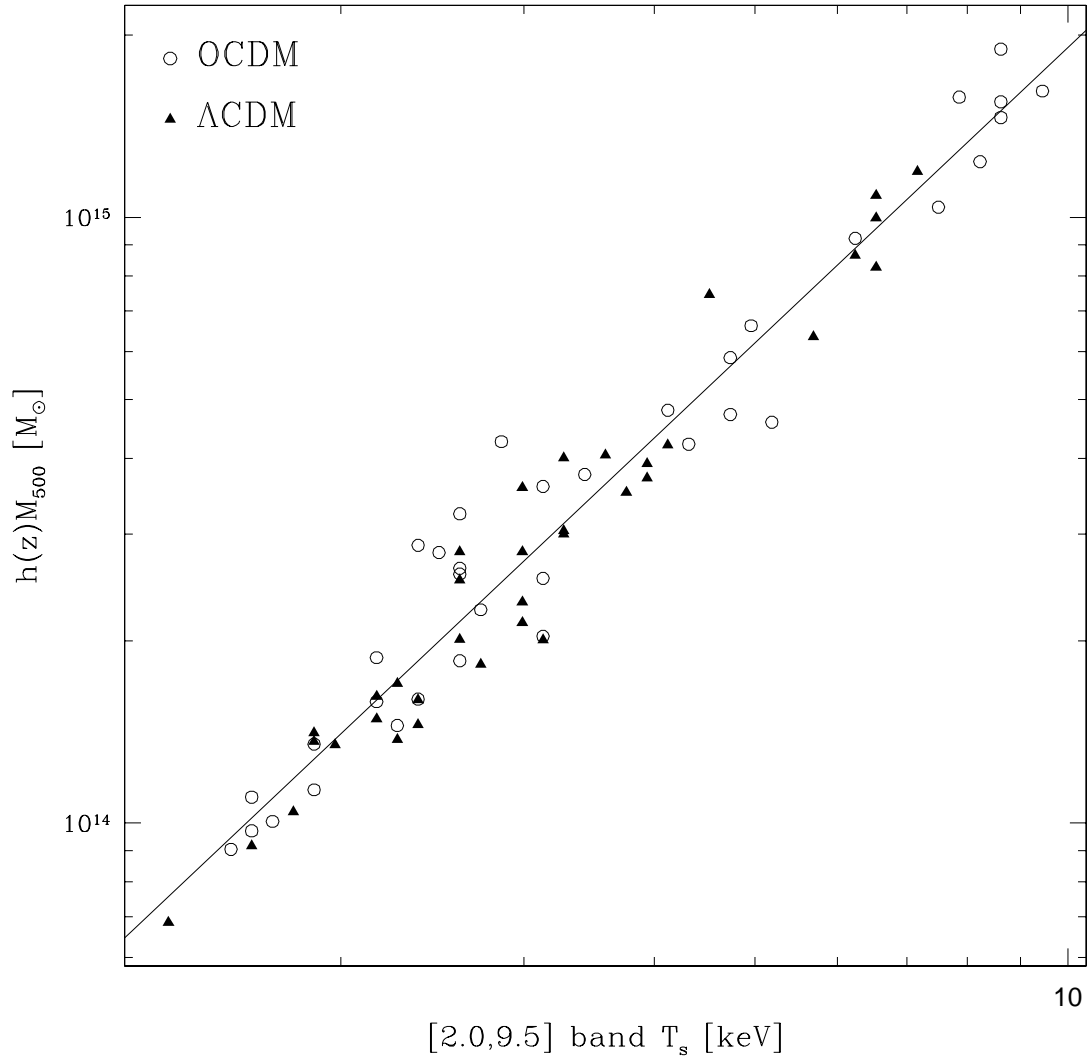}
\caption[The mass vs. spectral temperature relationship for simulated
clusters, measured within $r_{500}$]{Cluster mass vs. spectral
temperature for simulated clusters.  The solid line is the
least-squares fit to the data, and the corresponding equation is
given in Table \ref{mttable}. The scatter in $\log M$ at fixed $T_s$
is 0.081, or 20\%. Three output frames for each cluster are included
in the plot, corresponding to z = 0, 0.5, and 1. A similar result
with slightly larger scatter is obtained for measurements within
$r_{200}$.}
\label{specmasstemp}
\end{figure}

The scale dependence of deviations in the spectral temperature
will tilt the virial mass-temperature relationship. The relationship
between spectral temperature and total cluster mass derived from three
output epochs of the simulations is displayed in Figure
\ref{specmasstemp}.  This correlation has a significantly
higher slope and somewhat larger scatter than that displayed in
Figure \ref{virmasstemp}.  The best-fit $M_{\rm tot}$--$T$ relation
for these data has a slope of $1.62 \pm 0.07$ (90\% confidence level).
The baryon fractions of our simulated clusters within $r_{500}$
fall in the range $0.181 \pm 0.007$ (the global fraction is 0.2)
and are independent of mass, so for us $M_{\rm tot}$ and
$M_{\rm ICM}$ are essentially equivalent.
This result can therefore
be compared directly to the recent result of MME,
which measured the slope and scatter of the intracluster medium $M$--$T$
relationship in nearby clusters and found a slope of $1.98\pm 0.18$.
The degree of scatter in the two relations is similar; MME
found a scatter of 17\%, and the simulations display a scatter of 20\%.

The result displayed in Figure \ref{specmasstemp}
implies that some of the observed deviation from the virial relationship 
can be explained by invoking a normal incidence of both obvious and
hidden substructure, including both major mergers and the minor
accretion events which are typical of any epoch. Since both
Figure \ref{specmasstemp} and the result of MME depend on the
spectral temperature, their slopes can be compared directly.
Combining these two results, we can infer a mild 
temperature dependence in the ICM mass fraction of the form
$f_{\rm ICM} \propto T^{0.36\pm 0.19}$. This is wholly consistent
with the directly observed variation 
$f_{\rm ICM} \propto T^{0.34\pm 0.22}$ reported by MME. To put it another
way, the variation in $f_{\rm ICM}$ with temperature observed by
MME is not strong enough to account for the observed difference between
the $M_{\rm ICM}$--$T$ relation and the virial relation.  It is,
however, strong enough to match the $M_{\rm tot}$--$T_s$ relation
found in this paper.

We report the best-fit $M_{\rm tot}$--$T$ relationships from these
simulations for all our definitions of temperature in Table \ref{mttable}.
\begin{table}
\begin{center}
\begin{tabular}{lcccccc} \hline
Temperature & \multicolumn{3}{c}{$r_{500}$} &
	\multicolumn{3}{c}{$r_{200}$} \\
definition & $\log M_{15}$ & $a$ & $\sigma_M$ & $\log M_{15}$ &
	$a$ &  $\sigma_M$ \\ \hline \hline
$T_m$ (mass) & $-1.34\pm 0.02$ & $1.52\pm 0.03$ & 0.058 & 
	$-1.17\pm 0.03$ & $1.54\pm 0.05$ & 0.097 \\
$T_e$ (emission) & $-1.20\pm 0.03$ & $1.38\pm 0.05$ & 0.10 &
	$-1.05\pm 0.04$ & $1.39\pm 0.07$ & 0.15 \\
$T_s$ (0.5-9.5 keV) & $-1.21\pm 0.04$ & $1.48\pm 0.07$ & 0.14 &
	$-1.04\pm 0.04$ & $1.51\pm 0.08$ & 0.15 \\
$T_s$ (2.0-9.5 keV) & $-1.34\pm 0.02$ & $1.62\pm 0.04$ & 0.081 &
	$-1.19\pm 0.03$ & $1.64\pm 0.06$ & 0.10 \\ \hline
\end{tabular}
\end{center}
\label{mttable}
\caption{Mass-temperature relationships, defined as
$\log M_{\rm tot} = \log M_{15}+a\log kT {\rm [keV]}$, for all
definitions of temperature used in this paper.  The parameters
$M_{15}$ (mass in units of $10^{15} M_\odot$) and $a$ are calcuated
from the linear least-squares equation based on the scatter $\sigma_M$
in $\log M$ around the relation. The uncertainties represent a
one-sigma confidence interval.}
\end{table}
While the mass-weighted and [0.5,9.5] keV band spectral temperatures
are consistent with the virial relation, the [2.0,9.5] keV band
and emission-weighted temperatures are not. It is important to
emphasize that the emission-weighted temperature used in this paper
is not an observable quantity, and there is no evidence to suggest
that it is comparable to the observer's definition.  Further work with
these simulations needs to be done to reproduce an observable version,
for example the flux-weighted average of spectral temperatures
observed in each image pixel.

This analysis seems to imply that the best way to measure
temperatures with {\em Chandra} is to limit spectral fits
to photons of 2 keV or greater energy; the temperature
that results will not only be closer to the mass-weighted
temperature, but will also be comparable to previous determinations.
There is a price to pay, however, for rejecting the abundant low-energy
photons.  As we will see in the next section, the simulations
lead us to expect the presence of very cool gas not just in
cooling flows but also in the outskirts of the cluster. It is
well known that clusters are continually accreting smaller
concentrations of matter. Since they are less massive, the gas
in these subsystems is naturally much cooler than that in the
cluster. These clumps of remain visible in our simulated spectral
temperature maps for up to two billion years after the virial radii
of the two systems intersect. Including the soft photons decreases
the accuracy of spectral temperatures, but also increases the contrast
between these hot and cold regions. Thus, spectral temperatures in the
full {\em Chandra} bandpass will be more useful in determining a
cluster's dynamic state.

\section{Dynamics}

It is logical to ask several questions at this point.  First,
there are several instances where the spectral temperature
underestimates the mass-weighted temperature by a considerable
degree. If these excursions are the result of cluster dynamics,
we should be able to see evidence of unusual structure in the
cluster at these points.  Second, we have claimed that the
reason spectral temperatures typically underestimate the
mass-weighted temperature is that most clusters contain
significant amounts of cold gas in their outskirts.  If
accretion events and mergers are driving the difference
between the two temperatures, we should be able to discover
a time dependence in the effect, and perhaps even relate the
scale of the errors to typical accretion rates. Both of these
questions are really asking the question named in the title of
this paper: how are temperature measurements affected by a
cluster's dynamical state? 

Fortunately, the history of the simulations gives us a means
to answer this question.  As stated in \S 2, we have created
spectral temperature maps for each cluster at 16 points during
its evolutionary history.  We have also kept track of the
trajectories of dark matter groups during this evolution
and used this information to determine the times at which
major merger events occur.  Using this information, we can
create objective definitions of the cluster's dynamic age
and attempt to correlate these quantities with the biases
described in the previous section.

\subsection{Identifying Mergers}

The mergers themselves are easy to find. At each timestep
we identify dark matter groups with a mass of at least 100
particles (the real mass corresponding to this threshold
varies but is always a fixed fraction of the total mass in
the simulation volume). By keeping track of the particle
indices associated with each group, we can identify at later
times which groups were absorbed into a given cluster.
We identify merging groups with a mass ratio of at least
10\% as ``major mergers'' and build up a catalogue of these
events.  This results in a catalogue of 75 merger events for the
24 clusters. The limiting mass ratio was chosen to strike a balance
between frequency and significance; a merging subclump of 10\%
the primary mass can be easily seen in our temperature maps
(when the viewing angle is right) as a region of cool emission
behind a bow shock, but is not so common that simultaneous
events are typical. A fine example, which also happens to be
one of the two exceptions to this rule, is shown in Table
\ref{3waytmap}, which shows the spectral temperature map of a
three-way merger in the two bandpasses discussed earlier.
Although the merging mass ratio is small, it is clear in this
illustration that soft photons from the merging subclumps 
dominate the spectrum over a fairly large volume.  A minor
shock front is  visible between the two accreting systems,
but these few pixels have very little effect on the overall
temperature of the cluster.


\begin{table}
\begin{center}
\begin{tabular}{cc}
\multicolumn{2}{c}{surface brightness (log scale)} \\
\includegraphics{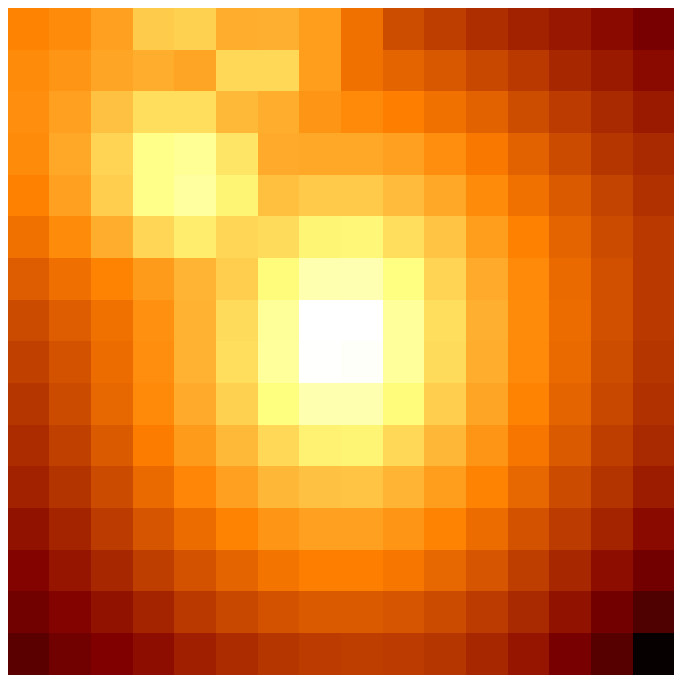} & 
\includegraphics{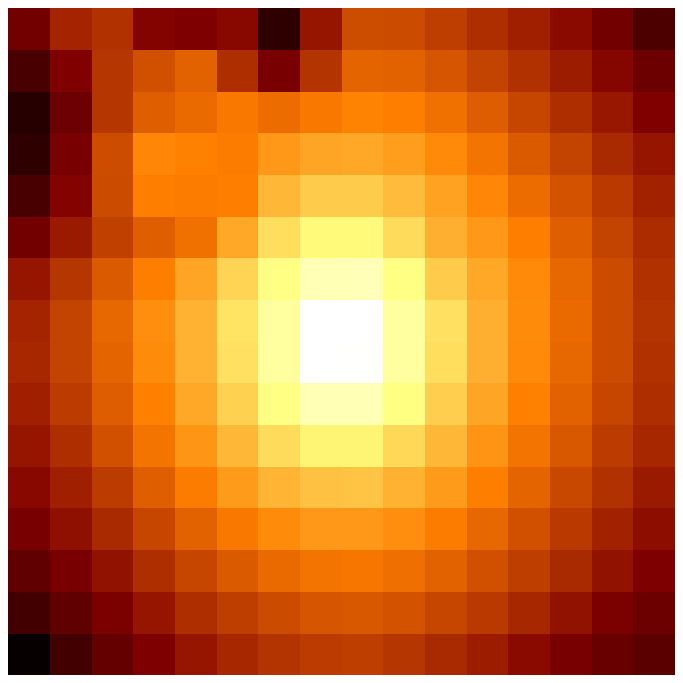} \\
\multicolumn{2}{c}{spectral temperature $T_s$} \\
\includegraphics{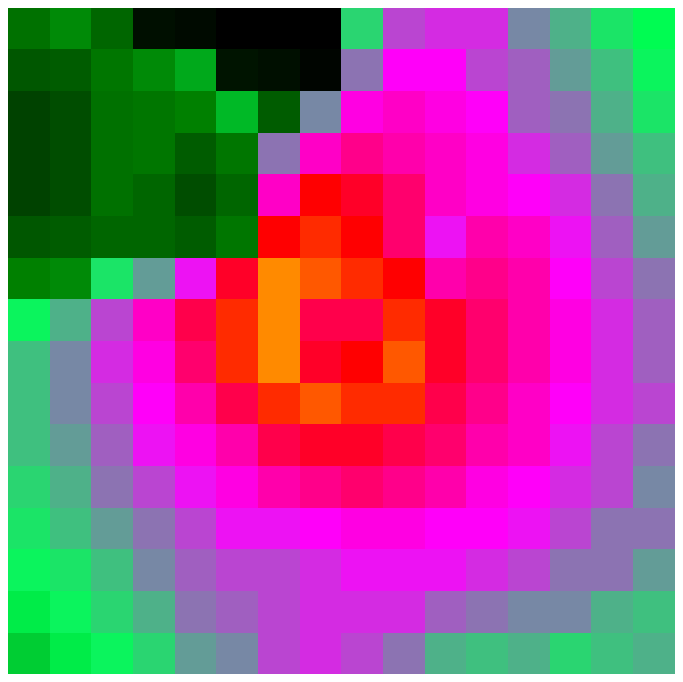} & 
\includegraphics{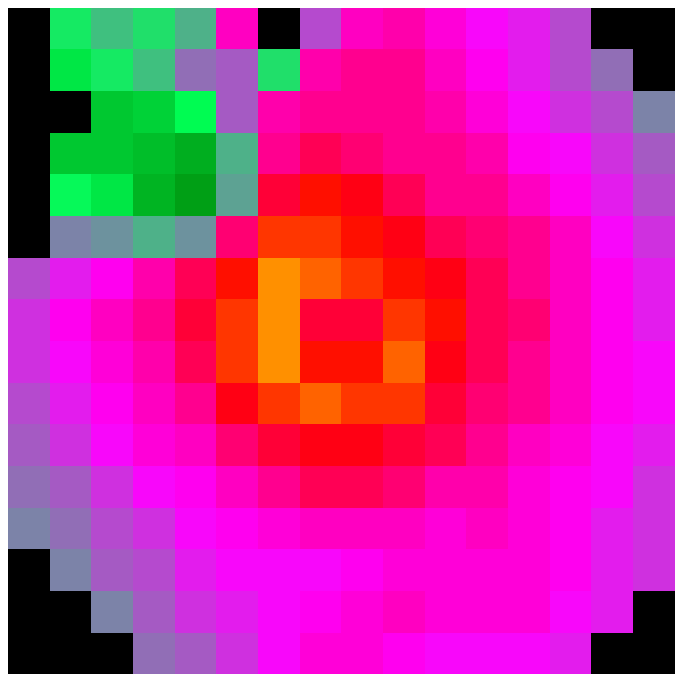} \\
0.25 keV & 4.0 keV \\
\multicolumn{2}{c}{\includegraphics{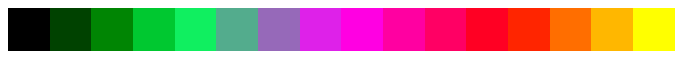}} \\
\end{tabular}
\end{center}
\caption{We see here one of the smaller clusters in our
sample as it goes through a three-way merger.  The larger
clump is 10\% the mass of the parent cluster, and the smaller
about 8\%. In the [0.5,9.5] bandpass (left-hand side)
the individual clumps are clearly distinguishable in the
surface brightness image. In the [2.0,9.5] bandpass (right-hand side),
however, this is not the case. In both bands, a shock can be seen
forming between the subclumps in the spectral temperature map.
The black pixels in the [2.0,9.5] keV $T_s$ map contain too few
photons to constrain the temperature. These images have a width
of $2r_{500}$, equivalent to 1.3 Mpc for this cluster.}
\label{3waytmap}
\end{table}

The time at which a merger occurs is a more nebulous concept
than the merger's existence. The individual outputs are separated
by about 0.6 Gy, so merely identifying this time with the first output
in which two groups are merged introduces a large random error.
We can, however, calculate the center of mass of each dark matter group in
previous outputs and create an interpolating function for
the groups' separation. This function can be used to create more
accurate and objective definitions of the merger time.

Our intuitive expectation for the effect of a merger
event on the ICM temperature is as follows: as the two groups
approach one another, their ICMs begin to interact at a fairly
large separation and we will see a shock begin to form.
The gas in a typical shock is much hotter than the rest of the ICM,
and should bias measurements towards higher temperatures.
As the event proceeds, the main
body of the merging system penetrates the virial radius of
the cluster and introduces a large mass of cold gas into the
image. The subclump will be more massive and more luminous than
the gas in the shock, so its weight will be greater under any
reasonable definition of the temperature.  We thus expect to
see a rapid cooling in the combined spectrum shortly after the
shock forms, even though the shock may still be clearly visible
in temperature maps. After some time the ICM will equilibrate
and approach the virial temperature again.  Most merger events
involve very small mass fractions, so this process should not
greatly affect mass-weighted temperatures.  We can hope to observe
the process of heating and cooling through a merger much more
clearly in the spectral and emission-weighted measures.

Practically, we interpret this expectation by identifying
a merger time $t_m$ for each event, defined as the time at
which the virial radii ($r_{200}$) of the two groups intersect
for the first time. Identifying the merger time in this
manner corresponds to the requirement that the mergers
have a uniform free-fall timescale, coming together in a
constant fraction of the Hubble time $t_H\equiv H_0(z)^{-1}$.
Further analysis of the dark matter halo dynamics reveals that the
centers of the two groups typically coincide about $0.2t_H$ after
$t_m$.

Another relevant time scale in this problem is the time required
for the ICM to approach hydrostatic equilibrium after
a merger. The equilibration time is often approximated by
the time  required for a sound wave to traverse the cluster,
\begin{equation}
t_s = 2\times 10^9 (kT)^{-1/2}D\ {\mathrm Gy}\ {\mathrm Mpc}^{-1}
\label{tcross}
\end{equation}
where $kT$ is
measured in keV and D is the diameter of the 
cluster (Sarazin 1986). The clusters in our ensemble have sound
crossing times between 1 and 4 Gy, with a mean of 2.3 Gy. The mean
time between merger events, on the other hand, is about 2.7 Gy. It
would seem from this that our clusters are out of hydrostatic
equilibrium for most of their total lifespan.

\subsection{The Effect of Mergers}

Armed with a list of 75 merger times $t_m$,
the Hubble time $t_H$ at each output frame, and the
sound crossing time $t_s$ in each output frame, we
can proceed to look for correlations of temperature
with the cluster dynamical states.  We start by defining
a parameter which measures the cluster's proximity to
a major merger:
\begin{equation}
\Delta\tau_H \equiv (t-t_m)/t_H
\end{equation}
where for each output frame $t_m$ is taken to be the time
of the {\em closest} merger event.  $\Delta\tau_H$ can
thus be either positive or negative, a freedom which
reflects our uncertainly as to when the merger process
will begin to influence the spectrum.

\begin{figure}
\epsfxsize 6.0in
\epsfysize 6.0in
\epsfbox{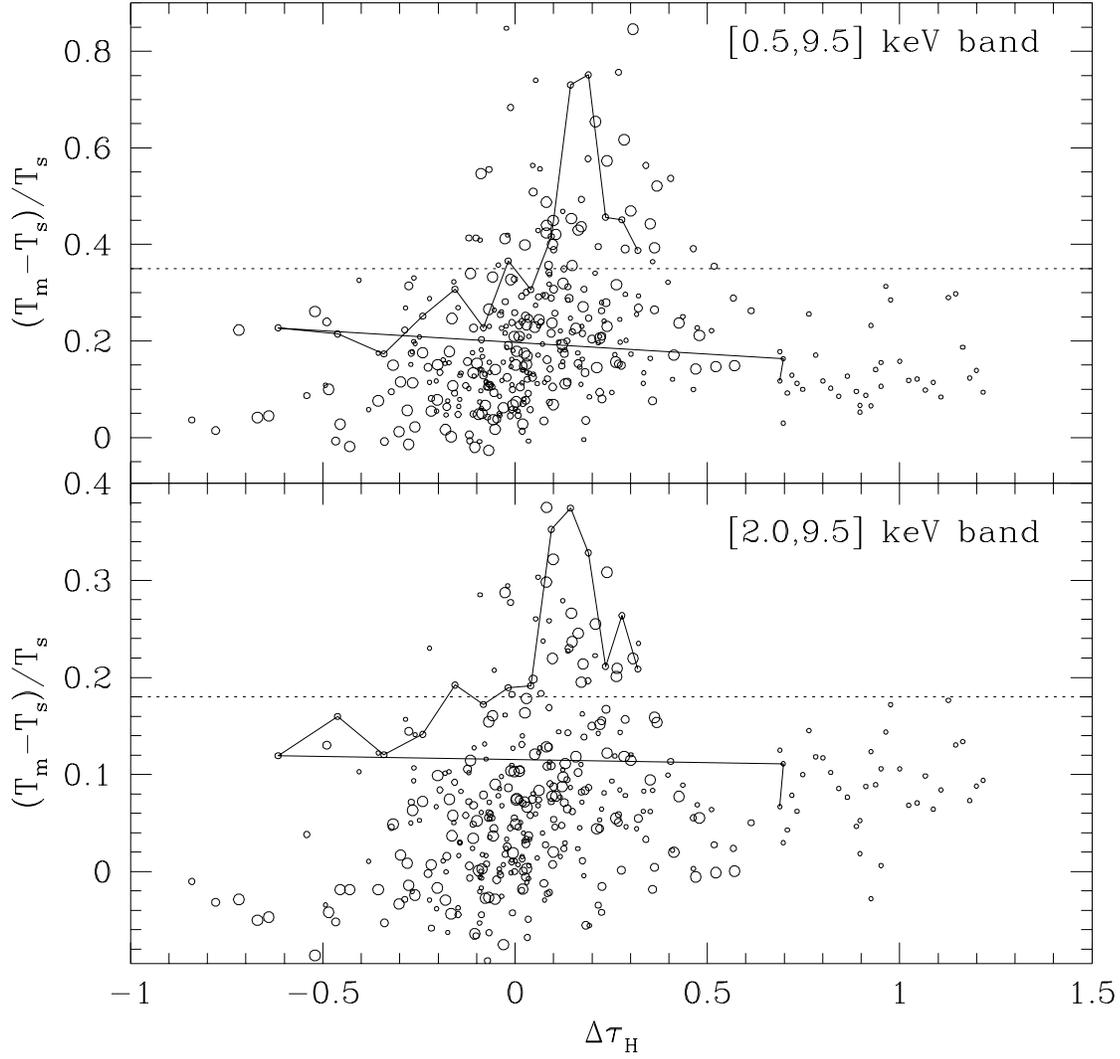}
\caption[The error in spectral temperature, $(T_m-T_s)/T_s$,
plotted against proximity to a major merger]{The fractional
deviation in spectral temperature as a function of $\Delta\tau_H$.
This figure and the variables displayed are described in
detail in the text. The point sizes increase with
the mass ratio of the closest merger event, and the
connected points correspond to the only cluster in our sample
which goes through just one merger in its lifetime. The
horizontal dotted lines are visual aids to help separate
the regions of large errors close to a merger event.}
\label{dtvst10}
\end{figure}

The relationship between spectral temperature errors
and $\Delta\tau_H$ is striking, and is displayed
in Figure \ref{dtvst10}. This Figure is constructed for
clusters observed in the $r_{500}$ window; the
corresponding Figure for $r_{200}$ looks similar.
The four different point sizes correspond to categories
in the merger mass ratio: from smallest to largest they
represent the ranges $m/M < 0.2, 0.2 \leq m/M < 0.4,
0.4 \leq m/M < 0.6$, and $m/M \geq 0.6$. A glance at
this plot demonstrates that the mass ratio of the merger
doesn't have much impact on the spectral temperature
measurements. This can be understood as evidence for a
balance in the roles of the subclump's temperature and
luminosity in causing the bias; in a major merger event
the subclump has more influence on the spectrum but
its temperature is also closer to the parent system's.
The fact that all of the large deviations are clustered near
the $\Delta\tau_H=0$ axis indicates that these excursions
are indeed related to the merger process, and suggests a method
for culling dynamically young clusters from our sample which we
will return to later. The horizontal line is a visual
aid to the scale of error which may be isolated from the
sample, drawn so that all the points above it are isolated
within a small range in $\Delta\tau_H$.

There is a single cluster in our sample which goes through
one and only one major merger in its lifetime; its points
are connected with a solid line in Figure \ref{dtvst10}
to serve as a benchmark for our physical intuition. Unfortunately,
this cluster is not entirely typical; it is rather small
and somewhat cool for its mass ($5 \times 10^{14} M_\odot$,
$kT_m =$ 4 keV, $kT_s$ = 2.8 keV). Still, the process of
apparent cooling as it goes through a merger with mass
ratio $m/M = 0.38$ is rather obvious.

We can reiterate this point by defining a second
dynamical time $\Delta\tau_s$ with respect to
the sound crossing time of the cluster:
\begin{equation}
\Delta\tau_s \equiv (t-t_m)/t_s.
\end{equation}
Although this equation looks very similar to 
our definition of $\Delta\tau_H$, there is
an important difference: we do not allow
this timescale to take on negative values.
Thus the merger time used is now {\em always}
the last merger to occur before the output
time $t$. For outputs where no merger has yet
occurred, we take the numerator to be the
age of the cluster. We define the time
$\Delta\tau_s$ in this manner because we
are interested in time required for our
clusters to come to equilibrium after a merger.

\begin{figure}
\epsfxsize 6.0in
\epsfysize 6.0in
\epsfbox{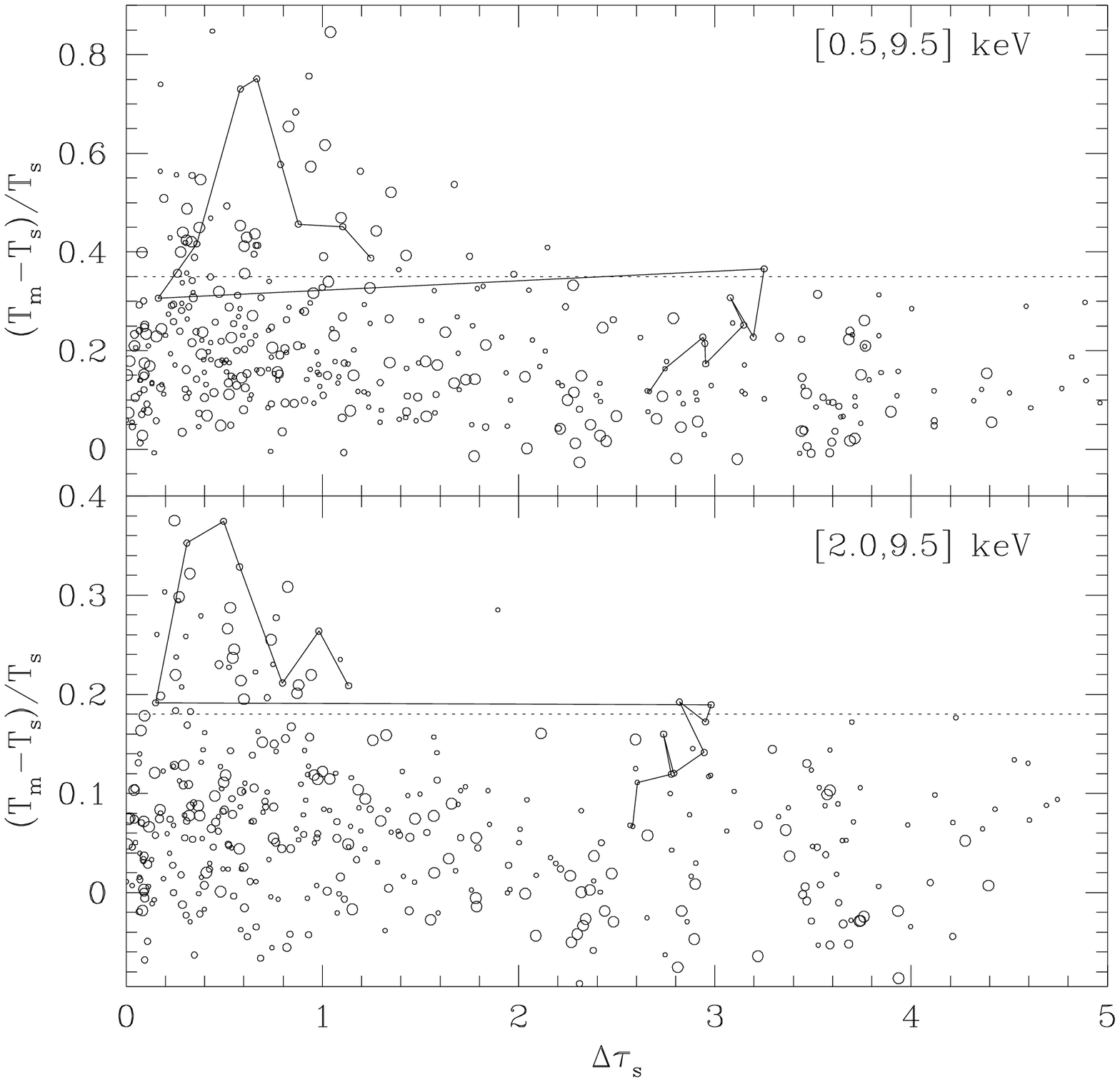}
\caption[The error in spectral temperature plotted against
$\Delta\tau_s$, the time elapsed since the last major merger
divided by the sound crossing time.]{The fractional deviation in
spectral temperature is plotted against $\Delta\tau_s$, as
described in the text. Most of the large differences occur in
frames less than one sound crossing time after a merger event,
especially in the [2.0,9.5] keV bandpass. These temperatures
are also measured within $r_{500}$.}
\label{dtvsteq}
\end{figure}

We plot the bias in spectral temperature against
$\Delta\tau_s$ in Figure \ref{dtvsteq}. The result
is similar in that the large errors mostly occur
close to a merger event and typically less than
one sound crossing time distant.  This suggests that
the measure given in equation \ref{tcross} is not
just a sound dimensional estimate but actually a
very good measure of the equilibration timescale.
Again, the results for other bandpasses and density
contrasts look quite similar.

It is interesting to note from Figures \ref{dtvsts}
and \ref{dtvst10} that the clusters which have large
temperature differences are different for the two bandpasses.
In the [0.5,9.5] keV range, the spectral temperatures
are more likely to carry large errors in the low-temperature
clusters.  In the [2.0,9.5] bandpass, we see the
largest errors associated with the hottest clusters.
Despite this difference, however, we see in Figure
\ref{dtvst10} and its equivalents that the largest errors
in both bandpasses are associated with merging clusters, and
moreover are concentrated in an interval about $0.2 t_H$ wide
after $\Delta\tau_H = 0$.

This tells us that merging clusters, and only merging clusters,
are likely to carry large temperature errors, but the different
bandpasses select clusters in different mass ranges.
In the [2.0,9.5] band the fit is driven almost entirely
by the effective slope of the bremsstrahlung component. This
slope varies rapidly at low temperatures, and the spectrum
of a blob of colder gas will be swamped by that of the
parent. At higher temperatures (for example, a 9 keV parent
and a 5 keV satellite) the mixing of exponential tails is
more likely to produce a significantly cooler spectrum.
In the [0.5,9.5] keV band, however, there is more
balance between the soft and hard photons. Between 0.5 and 2.0
keV the spectrum will be driven by the merging subclump,
while in the brehmsstrahlung tail the spectrum will be driven
by the parent cluster.  The tug-of-war between the two regimes
is decided by the number of bins in the spectrum, so in this
bandpass the hottest clusters are better able to resist the
influence of a cool component.

\subsection{Culling Mergers}

We have seen in the previous sections that a simulated cluster
ensemble, with realistic incidence of substructure, displays
spectral temperatures with a significant and scale-dependent
systematic bias. The largest deviations between spectral and
mass-weighted temperatures come from clusters
which are dynamically young: they are either about to undergo or
have just undergone a major merger and the presence of cool
gas in the observation window changes the shape of the combined
spectrum. Some of these mergers will be easily identifiable with
{\em Chandra} through the temperature structure of the gas, displaying
images similar to those displayed in Table \ref{3waytmap}.
Others will be invisible, however, hidden behind the main body
of the cluster but still significantly altering the spectra.
We would like to develop a method for identifying some or all
of these, so that mass-weighted temperatures and cluster binding
masses can be determined through X--ray observations with a
minimum of error.  Our expectation is that a large sample of
cluster observations, culled of all of obvious mergers and some
inobvious mergers, will reflect the virial relations better
and allow us to more accurately measure variations in baryon
fraction, deviations from self-similarity, and constraints on the
power spectrum.

The results of the previous section tell us that the clusters
with spectral temperatures much colder than the mass-weighted
temperature always lie close to a merger, although the reverse
is certainly not true.  We are left with little else to go
on: the deviation in spectral temperature is uncorrelated with
either merger mass fraction (a bit suprising) and shows no
dependence on the cosmological model. It does
show some dependence on the mass of the parent, which leads
to a change in slope of the observed mass-temperature
relation. Figures \ref{dtvst10} and \ref{dtvsteq} of the
previous section are not observables, since we can't yet
reliably identify merger events. We would like to cull the
largest excursions from a real cluster sample, but we need to
do it some other way.

The solution is to build up an observable mass-temperature
correlation (or ICM mass, if you prefer; remember that the
fixed baryon fraction in our simulations makes the two equivalent)
and cull out the clusters which are coolest {\em relative to the
mean relationship}. We have already plotted the cluster mass vs.
spectral temperature for the [2.0,9.5] bandpass within $r_{500}$
in Figure \ref{specmasstemp}. Our analysis leads us to expect
that the clusters which fall far below the best-fit line
may be merging clusters.  To test this we treat mass as
the independent variable, and calculate the difference between
$T_s$ as measured and the predicted $T_s$ based on the best-fit
correlation with cluster mass.  This quantity is then
plotted against $\Delta\tau_H$, as defined in the previous section.

\begin{figure}
\epsfxsize 6.0in
\epsfysize 6.0in
\epsfbox{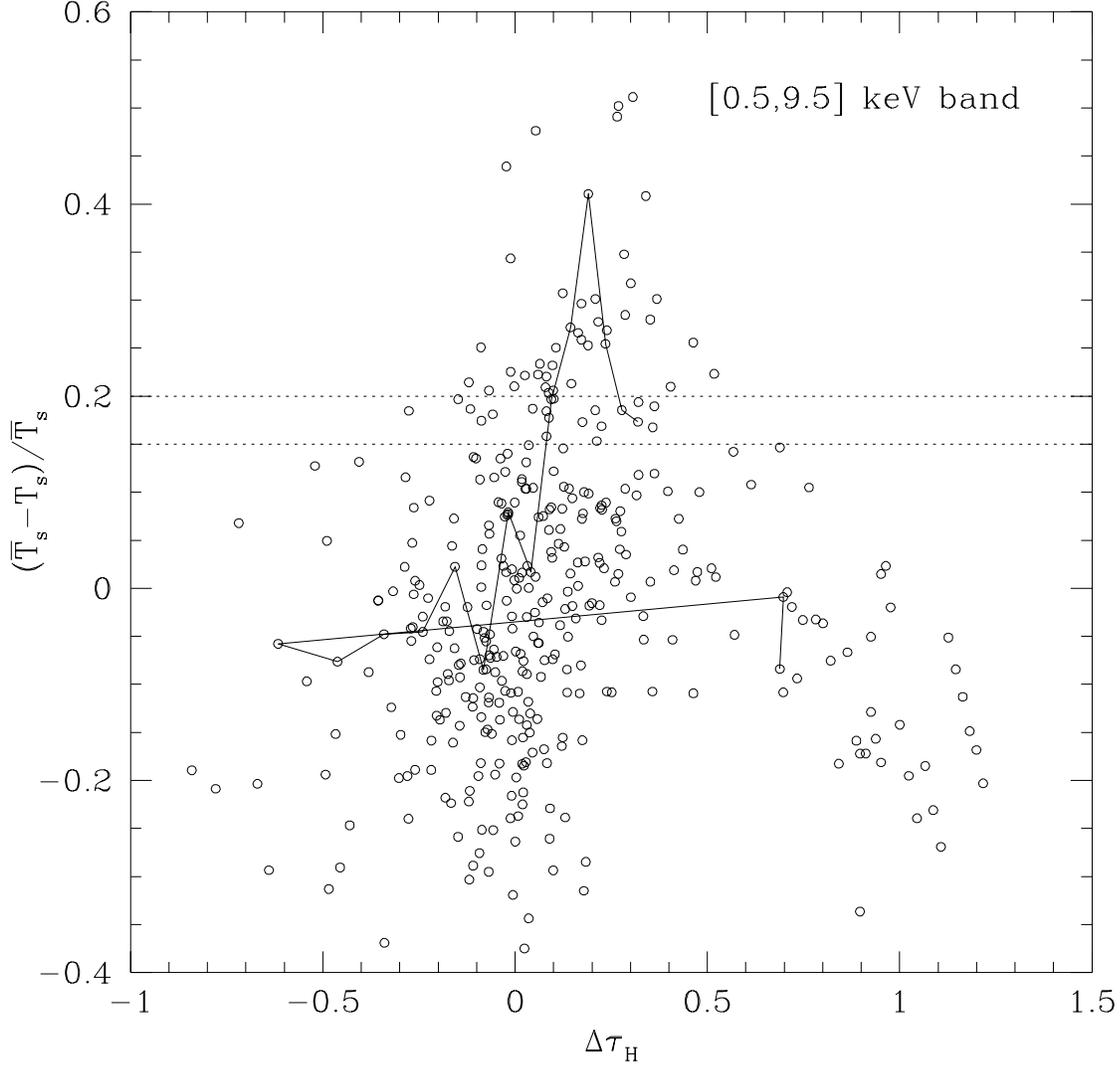}
\caption[The difference between spectral temperature and
the mean $M$--$T_s$ relation as a function of dynamical
time $\Delta\tau_H$.]{The fractional difference between $T_s$
and $\bar{T}_s$, the spectral temperature expected from
the best-fit line for that mass, is plotted against dynamical
time. Again, the clusters which are coolest relative to
the mean relation are clustered in the approximate range
$-0.1 <\Delta\tau_H < 0.4$. This plot implies that merging
clusters can be culled from a sample in a statistically
well-defined way by observing the mass-temperature relation
and making a similar cut on the data.  The spectral
temperatures used here are measured within $r_{500}$
in the [0.5,9.5] bandpass.}
\label{dtmvst10}
\end{figure}

The results for the [0.5,9.5] bandpass within $r_{500}$ are
displayed in Figure \ref{dtmvst10}.  We've switched
bands here because the culling technique we propose seems to
work best when temperatures are measured over the entire
energy range (due to the increased contrast mentioned above).
This plot is harder to interpret, but by drawing a
horizontal line we see that we can isolate the 40 (10\%)
coolest (relative to the expected temperature at that mass)
images and be reasonably sure that these clusters are
dynamically young, i.e. $-0.1 \lesssim \Delta\tau_H \lesssim 0.4$.
In real life, we could apply this technique by (1) measuring ICM gas
masses with precision in a soft bandpass such as that
of the {\em ROSAT} PSPC where the luminosity is a weak
function of temperature (e.g. Mohr \etal 1999); (2)
measuring spectral temperatures with {\em Chandra}; and
(3) culling clusters from the sample which deviate
by more than $-20\%$ from the mean relation.  Needless
to say, this would work best with a large survey and
would require a large time investment--but the remainder
of the sample would suffer less bias from clusters which
are dynamically young. 

Note that the definition of ``dynamical youth'' defined above
is empirically driven.  At low redshifts, 40\% of
the Hubble time is a generous 3-5 billion years, typically
about twice as long as the freefall ($\sim 0.2t_H$) and sound
crossing timescales. We adopt this definition solely because
all of the clusters with large temperature errors fall within
this range, without further justification, since our main
interest lies in removing the least accurate measurements from
our sample. Out of the whole ensemble of 384 images, fully 70\%
lie within this range of dynamical states.

It is also interesting to note that the hottest spectra
occur either long after a merger event or at times
$\Delta\tau_H \lesssim 0.2$. The free fall time from
$r_{200}$ is about $0.2t_H$, so the time interval from
$\Delta\tau_H = 0$ to $\Delta\tau_H = 0.2$ represents the
time required for a merging system to completely enter the
observation window.  The equivalent plot for $\Delta\tau_s$
is not given here, but again shows that most of the clusters below
this cut are dynamically young. ($\Delta\tau_s \lesssim 1$).
A similar result is found in constructing these plots
with the emission-weighted temperature, except there is
also mild evidence for shock heating in temperatures
determined previous to $\Delta\tau_H = 0$.

\begin{figure}
\epsfxsize 6.0in
\epsfysize 6.0in
\epsfbox{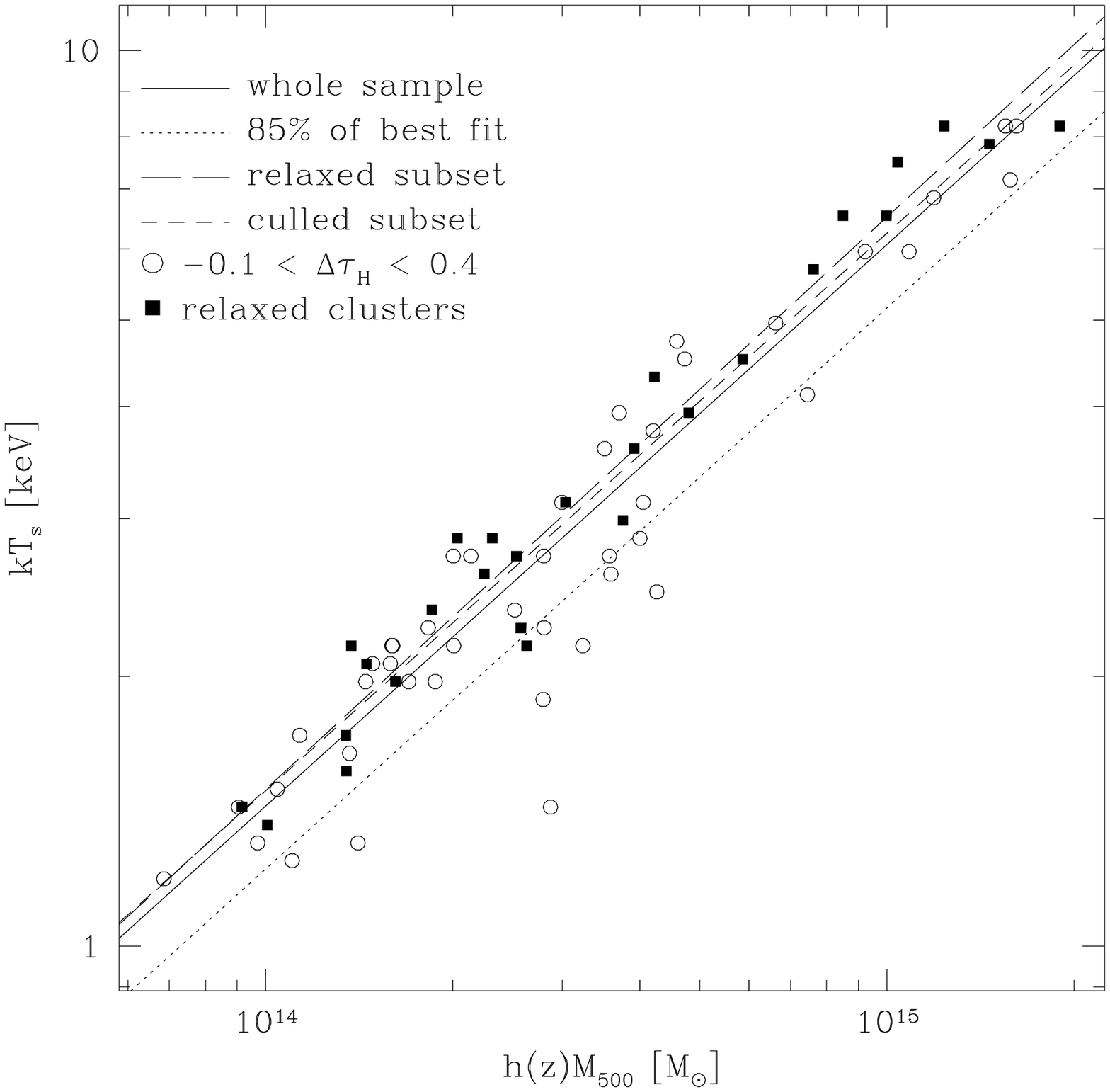}
\caption[The culling method of Figure \ref{dtmvst10} is
demonstrated on the three-redshift subset of images.]
{The culling method implied by Figure \ref{dtmvst10} and
described in the test is demonstrated on a subset of
images corresponding to redshifts 0, 0.5, and 1.0. The solid
line is a fit to the entire sample (as reported in Table \ref{mttable});
the dotted line has the same slope but a temperature
normalization 15
easy to see from this plot that most of the clusters
below the best-fit line lie close to a merger event,
and performing a cut on clusters which are cool relative
to the fit preferentially selects for such objects. Also
plotted are linear fits to the subset of ``relaxed''
clusters (black squares), and the subset of clusters
above the dotted culling threshold.}
\label{cullmasstemp}
\end{figure}

Finally, we apply this analysis to the three-redshift
subset of our images and their mass-temperature relationship.
Figure \ref{cullmasstemp} plots spectral temperature against
total mass in the [0.5,9.5] band within $r_{500}$, draws the
best-fit line and a cut on clusters 15\% cooler than this line,
and identifies those clusters which are dynamically young
with a different symbol.  We can see from this plot that
the proposed cut preferentially selects out merging clusters,
although there are still many left in the sample.
A previous analysis of these simulations (Mathiesen \etal 1999)
determined that ICM masses of regular clusters are measurable with
only $\sim 5\%$ scatter and $\sim 10\%$ bias using the parametric
technique of Mohr \etal (1999), and more importantly that the
degree of error in $M_{\rm ICM}$ is uncorrelated with mass and
temperature. This culling method would therefore be unchanged if
we attempted to apply it to the observable $M_{\rm ICM}$--$T_s$
relationship.

This cut does not change the slope
of the observed $M$--$T_s$ relation and has a small (about 0.5
sigma) effect on the normalization. This method is thus primarily
useful for identifying a subset of dynamically
young clusters in a sample, presumably in
the hopes that some will be found which aren't obvious in other
ways. We have selected out ten merging clusters and a single relaxed
object; analysis of Figure \ref{dtmvst10} implies that this ratio would
be similar in a larger sample. 

\section{Conclusions}

The first (and perhaps the most suprising) result
of this paper is the revelation that the emission spectrum
of a realistically complex intracluster medium is
barely distinguishable from that of an isothermal gas, even
over a broad spectrum.
This correspondence arises because we have created
spectra with rather coarse (150 eV) bins in an effort to
match our observation model to {\em Chandra's} ACIS
instrument.  We provide relationships between spectral, emission,
and mass-weighted temperatures in two different bandpasses
and within two density constrasts; in particular, we find
that spectral temperatures determined in the [2.0,9.5]
keV range within $r_{500}$ are very similar in nature to
those determined by the {\em Einstein} MPC, {\em Ginga}, and
{\em EXOSAT}. The other bandpass, [0.5,9.5] keV, is designed to
be similar to {\em Chandra} observations, and can be used to
better interpret its results.

We find that realistically determined spectral temperatures are
commonly 10-20\% lower than the mass-weighted temperatures
in these simulations, a fact which has important implications
for cluster physics.  The bias arises through the natural and
frequent occurence of minor accretion events: small clumps of
cool gas which merge into the ICM and produce an excess
of soft X-rays, biasing the spectrum towards cooler temperatures
until they are fully assimilated. Because the mass-weighted
temperature follows the virial relationship, it is a more accurate
indicator of the binding mass.
Previous measurements of the cluster mass function
power spectrum need to account for this source of
error if they make use of spectral temperature data. We have
calibrated the relationship between spectral and mass-weighted
temperature for the [2.0,9.5] keV bandpass within $r_{500}$
for this purpose; this should allow an appropriate correction
for clusters observed by satellites with a similar energy range.

The scale-dependent nature of this bias changes the slope of the
observed $M_{\rm ICM}$--$T_s$ relationship in a direction consistent
with recent determinations, but it does not account for the whole
difference. Additional physics, such as a variation in the ICM
mass fraction with temperature, is still needed to explain the
observed slope.

Although we are stuck with a discrepance between observed
temperatures and mass-weighted temperatures, this bias can be
useful in identifying clusters which are dynamically young.
Not all of these events will be obvious, even
with {\em Chandra's} high-resolution surface brightness
and temperature maps; some of the mergers will be occuring on
an axis near our line-of-sight. In such cases the shocks and cool
regions will probably be masked by core emission, but the presence
of a cool subclump can still produce an unusually large deviation
in the spectral temperature.  We should not fall into the trap
of assuming that a cluster is relaxed because it looks spherically
symmetric; rather, we should do everthing we can to determine
whether or not it is truly relaxed. The simulations
show that all of the clusters with temperatures falling
far ($\gtrsim 15\%$) below the mean mass-temperature relation
are dynamically young; a subset of merging clusters can
thus be identified as outliers in the observed relation.

The results described in this paper are a necessary step towards
accurately measuring cluster masses under the assumption of virial
equilibrium, as well as towards a reliable technique for determining
a cluster's dynamical state observationally. It is not, however,
the only step: first the astronomical community must recognize
that the various definitions of temperature in common use are
{\em not} equivalent. Further work needs to be done in modeling
observed temperatures (e.g. core-excised spectral temperatures and
flux-weighted temperatures averaged over an image) and the effects
of substructure on observable quantities. Spectral modeling of
simulations is computationally more expensive, but not difficult,
and the level of systematic error caused by inaccurate modeling
is now comparable to the observational constraints on cosmological
parameters.

The spectral cubes described in this paper will be made avaiable
to the public on Mathiesen's research web site
\footnote{http://redshift.perseus.edu/bfm} along with
higher-resolution images for the subset of images used in
fitting the mass-temperature relationships, the accretion
history of each cluster, and documentation describing
the file format and how to model other satellite responses.
Our simulations of spatially resolved emission spectra
can be used to help interpret {\em Chandra} and {\em XMM}
observations of real clusters, and we are sure that enterprising
researchers can mine these datasets for other interesting
results.

\section{Acknowledgements}

We would like to thank Nancy Brickman, John Raymond, Randall Smith,
and Martin Sulkanen for their help in implementing the Raymond-Smith
and {\texttt mekal} spectral models. This research was supported
by NASA grants NAG5-2790, NAG5-7108, and NAG5-8458; and NSF
grant AST-9803199.

\end{document}